\documentclass[8pt,a4paper]{book}

\usepackage{amsmath, amssymb, amsfonts, latexsym , mathabx}
\usepackage[utf8]{inputenc}

\usepackage{graphicx}
\usepackage{anysize} %Partición de palabras al final de la linea

\usepackage{multirow} % para las tablas

\usepackage{hyperref} % Habilito el uso de referencias (accesos directos a ecuaciones y temas)

\usepackage[toc,page]{appendix} % Los apéndices

\usepackage{lmodern}
\usepackage[T1]{fontenc}
\usepackage{theorem}

\theoremstyle{break}

\marginsize{2.5 cm}{2.5 cm}{2 cm}{2 cm}

\title{ \textbf{ \texttt{Relationship between rotation curves and matter distribution in spiral galaxy discs} } }
\author{\Large{Author: Gonzalo Fernández-Torija Daza\footnote{Departamento de Física, Univ. Carlos III, Madrid, Spain. E-mail: gonferna@fis.uc3m.es}} \\ Supervisor: Martín López Corredoira\footnote{Instituto de Astrofísica de Canarias, La Laguna, Tenerife, Spain}}
\date{\today}

\newcommand{\diffI}[2]{\dfrac{d #1 }{d #2 }}

\newcommand{\pdiffI}[2]{\dfrac{\partial #1 }{\partial #2 }}

\newcommand{\spaceI}{\hspace*{0.25cm}}
\newcommand{\spaceII}{\hspace*{0.5cm}}

\newcommand{\abs}[1]{\left| #1 \right|}
\newcommand{\norm}[1]{\left\lVert #1 \right\rVert}
\newcommand{\convergence}[4]{ #1 \overset{ #3 \rightarrow #4 }{\xrightarrow{\hspace*{1cm}}} #2 }
\newcommand{\entonces}{\spaceI \Rightarrow \spaceI}
\begin{document}

%AJUSTES INICIALES

\begin{titlepage}				% Se puede omitir
\maketitle  % Se ejecuta la portada (IMPORTANTE)
\thispagestyle{empty} % para que no se numere esta página
\end{titlepage}				% Se puede omitir

% \newpage % Dejamos página en blanco
%\mbox{}

%El índice -----------------------------------------------------------------
\tableofcontents % indice de contenidos

%\cleardoublepage
%\addcontentsline{toc}{chapter}{Lista de figuras} % para que aparezca en el indice de contenidos
%\listoffigures % indice de figuras

%\cleardoublepage
%\addcontentsline{toc}{chapter}{Lista de tablas} % para que aparezca en el indice de contenidos
%\listoftables % indice de tablas

%%%%%%%%%%%%%%%%%%%%%%%%%%%%%%%%%%%%
%EMPEZAMOS A ESCRIBIR

% Si queremos prefacio (solo vale para artículos)---------------------------------------------
% \begin{preface}  \end{preface}

% Si queremos abstract (solo vale para book y report, ocupa una página)--------------------------------
% \begin{abstract}
 
% \end{abstract}

\subsubsection*{\centerline{Abstract}}
This master's thesis is based on J. Q. Feng \& C. F. Gallo's  2011 article \cite{FengGallo2011}. These authors developed a numerical method of deriving rotation curves from the density distribution and, in particular, the inverse problem while considering just a self-gravitating disc and the thin disc approximation. Our first aim here is to reproduce the same analysis and expand it with various ideas and examples. Moreover, we add a final chapter extending the problem and its method to a third dimension through the perpendicular to the galactic plane.

The main obstacles to building this numerical implementation are certain singularities that they solve by carrying transformations out in the original governing equations and their numerical representation. Even though we have obtained clear and credible results, we found various problems in the inversion involving the final linear system matrix resulting in various stability problems. We try to fix these instabilities using different methods.

The dark halo (whose density is usually represented by a nearby spherical distribution) is supposed to support the outer parts of the rotation curves of spiral galaxies. Here, however, we work only with a self-gravitation disc. To treat this topic we first calculate the disc density distribution from measured rotation curve data of the Milky Way. We then compare this distribution with the observed exponential stellar density, and the difference is attributed to a dark disc. This representation of Feng \& Gallo of the Galaxy with a dark disc instead of a dark halo is controversial.

When we analyse the effect of flares on rotation curves, the thin disc approximation fails, and we need to introduce a vertical dimension to measure and predict the effects of the flare through different heights. Just by spreading the mass perpendicularly to the plane -- without adding any further mass -- the flare provokes no severe changes on the rotation curve. The flare mainly provokes a faster velocity decrease in the outer part of the Galaxy ($r\gtrsim 14$ kpc). But we also have obtained a slight velocity increase in the first kiloparsecs after the starting point of the flare.

\subsubsection*{\centerline{Resumen}}

El orígen de este Trabajo de Fin de Master es el artículo publicado por J. Q. Feng y C. F. Gallo 2011, \cite{FengGallo2011}, en el cual se desarrolla un método numérico con el fin de calcular las curvas de rotación galácticas a partir de una distribución de densidad dada y (ante todo) del problema inverso. Se considera para ello un disco autogravitante y una aproximación de disco fino. Nuestro primer objetivo es reproducir el mismo análisis y expandirlo con algunas ideas y ejemplos. Aún así, añadimos un último capítulo extendiendo el problema a la tercera dimensión a lo largo del eje perpendicular al plano galáctico.

El principal obstáculo que nos hemos encontrado  para la implementación de este método numérico son algunas singularidades que ellos resuelven realizando transformaciones en las ecuaciones originales y su representación numérica. A pesar de haber obtenido resultados claramente verosímiles, hemos encontrado problemas en el problema inverso relativos a la matriz del sistema lineal que nos han llevado a algunos problemas de estabilidad en la resolución. Hemos tratado de paliar esta inestabilidad usando diferentes métodos.

El método que estamos describiendo contiene un significado físico claramente controvertido, dado que estamos trabajando sobre un disco autogravitante. El halo oscuro (cuya densidad es normalmente representada con una distribución cuasi-esférica) se supone que soporta la parte más externa de las curvas de rotación. Para tratar este tema, lo primero que hacemos es calcular la distribución de una curva de rotación dada de la Galaxia. Más tarde, construimos un modelo clásico de la Galaxia compuesto de bulbo+disco+halo y finalmente comparamos ambas distribuciones nominando ``disco oscuro'' a la parte restante. Obtenemos una densidad superficial del disco oscuro, obteniendo un valor claramente alto comprado con la usual representación bariónica de un disco exponencial.

Cuando analizamos el efecto del \textit{flare} galáctico  sobre las curvas de rotación la aproximación de disco fino falla y necesitamos introducir una dimensión vertical para así medir y predecir los efectos del \textit{flare} a diferentes alturas sobre el plano galáctico. Sin añadir cantidad de masa alguna, solamente repartiendo la masa a lo largo de la dirección perpendicular al plano, la curva de rotación no se ve sometida a notables cambios debido al \textit{flare}. Los cambios que se producen son principalmente un decrecimiento de la velocidad en las zonas exteriores de la galaxia ($r\gtrsim 14$ kpc), pero también se observa un tímido incremento de la velocidad en los primeros kiloparsecs justo donde el flare comienza.

% ==================================================================================================
% CHAPTER 1 ========================================================================================
% ==================================================================================================

\chapter{Introduction}

% Section 1.1 ======================================================

\section{Historical introduction to the disc problem}

The rotation curve of galaxies and its mass distribution as the main cause are the principal characteristics of this master's thesis. The problem is studied mainly in late type spiral galaxies, where the evidence is more clearly seen than in other stellar types. Such galaxies are more common in the low-density part of the universe ($\sim$80\%); in cluster cores its number falls notably ($\sim$10\%). Studying this topic will bring us to a better understanding of our Galaxy, the Milky Way, and its kinematics and dynamics. In recent years, a large number of stellar surveys of the Galaxy have been taking place and have culminated in the GAIA mission, which is now providing us with very accurate data from our galactic neighbourhood. However, other famous surveys, such as the Sloan Digital Sky Survey (SDSS), have also been gathering enough information on the stellar disc and its substructures to enable theoretical studies to reach various  results. The main idea of this dissertation is founded on the articles of \textit{James. Q. Feng \& C. F. Gallo} \cite{FengGallo2011,FengGallo2014,FengGalloDarkMatter} (hereafter, F\&G), which exemplify the approach adopted here.

To avoid tediously repeating the same references throughout this dissertation, we note here that most of the information used herein has been taken from references \cite{Binney}, \cite{Sofue2013}, \cite{NestiSalucci2013}, \cite{VanDerKruit1978} and \cite{VanDerKruitFreeman2010}.

\subsubsection*{Overall structure of the galaxy}
In the present literature we can easily find the basic structure of spiral galaxies, and with regard to our Galaxy, which seems to be a late type one. The bulge, disc, and halo are outline its fundamental structure, making the Milky Way an SBc type galaxy. However, we also know there are more substructures, such as spiral arms, the central bar, warps, flares, etc., that can be deduced from observations. These features can exert an important influence on the kinematics and star formation. They also can provide us with hints concerning the secular evolution of the Galaxy. But our topic concerns kinematics; each of the basic components dominates the dynamics of some specific radial section. The bulge produces a high peak in the first few hundreds of parsecs, followed by a steeply descending curve that has little influence on other radial sections. Further, the disc produces a gradual descent, an almost flat part where there could be some irregularities produced by some of the aforementioned substructures. Finally, the halo seems to dominate the outermost part and the outskirts of the Galaxy.

\subsubsection*{Bulge and bar}
Even though the bulge is not the main topic of this master's dissertation, there are certain points related to its structure that will need to be considered. Owing to the high extinction level, the mass distribution of the bulge remained largely unknown, especially in the innermost part ($<$ 2 kpc) until advances in infrared observing in the 1970s. At least two components have stood in the latest state-of-the-art studies: a more condensed bulge extending up to $\sim$ 1.5 kpc and a bar that could be extend out to $\sim$4 kpc. But in more detailed studies we can even consider the central massive black hole nesting in the centre and, in general, a massive core within the bulge. The contribution of the black hole is not so relevant in the scale plotted and it is usually treated as dominant object in the very central region ($\sim$1 pc, \cite{SofueCurves2013} ). The Galactic core will be represented by the parameter ($R_c$). We will have the chance of analysing its influence on the main method described in this essay, as we shall see in the second and third chapters. However, the velocity anisotropy limits the accuracy of the rotation velocities there. The evident uncertainties and non-circularities of the curves that we find at $\lesssim$2.5 kpc will prevent us from dedicating any special care of the rotation velocities in that region.

\subsubsection*{The disc and its substructures}

The observation of the galactic disc in a late-type spiral galaxy is clearly relevant when we wish to draw conclusions about its structure, substructures, dynamics, and even secular evolution. Activities arising from such processes and features as star formation, barred structures, or spiral arms have a clear influence and development in the disc. Concerning to our Galaxy, the first description of our stellar neighbourhood concerning the flattened nature this stellar structure was made by William Herschel ({\textit{On the Construction of the Heavens}, 1785). According to Herschel's model, the Sun was close to centre of the Milky Way. Modern studies concluded that Herschel only took into account stars with magnitudes close to $V \approx 15$. Herschel, along with many later investigators, was unaware of interstellar absorption. Kapteyn \& van Rhijn in 1920 and Kapteyn in 1922 included the reddening of the starlight with the distance, but interstellar absorption as such was not taken into account; existence of interstellar absorption was not discovered until 1930 by Trumpler. And the Sun in the Kapteyn model of the `Universe' was again placed close to the centre of the disc. Nevertheless, the flattened shape of the star distribution was confirmed. 

In the disc, most of the mass is represented by stars, although $\sim$10\% of the total mass seems to come from the interstellar medium. The thickness of the disc is limited to a few hundred parsecs. If we consider locations very far from the plane, the effects of the stellar mass on rotation curves are almost negligible, so the thin disc approximation is quite justified here. We will usually take a ratio of 0.01 between the disc and the maximum considered radius. Here, we understand thickness to be defined by the disc stars. The molecular clouds, HI gas, and even the brighter OB stars usually have a more bounded vertical domain ($\lesssim$ $100$ kpc). On the other hand, we can find hot gas out to $\sim$2 kpc from the plane.

Regarding kinematics, the first steps in our problem were taken in the first decades of the 20th century. The observations of M\,81 and M\,104 (objects considered nebulae on those times) by Wolf and Slipher in 1914 revealed inclination of the stellar absorption lines in the central regions. The identification of these nebulae as galactic discs was fairly polemical. Pease in 1916 and 1918 was the first to show a plot of a radius versus velocity of the same objects and his results, although they do not tally well with recent ones, led him to confirm the rotation of those nebulae. Later, in the 1950s, while Jan Hendrik Oort was observing the rotation velocities of the stars in the solar neighbourhood, he observed higher rotation curves compared with the ones predicted. Oort did some approximate calculations to get the galactic mass within the solar radius. Although this result was not so far from the modern value, he estimated it with the assumption of the galaxy as a sphere. But we know there is no such simple situation. The point here is that we must analyse the rotation curve and consider the force balance; a `struggle' between the centrifugal and the gravitation forces. And, as we will see in this essay, the disc attraction does not have the same structure as the spherically distributed gravitational influence. Oort also checked other galaxies and he found in galaxies such as NGC\,3115 a good example of how the light emitted bears no clear relation to galaxies' rotation curves and/or mass distribution. This storyline has its real origin in those decades. But it was not really considered until the decade of 1970. The technology had to be developed in the 1960s to be able to read H$\alpha$ and NII emission lines arising from HII regions in spiral galaxies. The first `spider diagram' was published by Argyle \cite{Argyle1965} in 1965, and the galaxy chosen was M\,31 (Andromeda). Many rotation curves were taken and many diagrams were published throughout those years.

The measurements of these galactic rotation curves of the Milky Way revealed the famous diagram of longitude--radial velocity (LV) diagrams taken along the Galactic plane. The usual representation comes from emission lines of HI ($\lambda$ = 21 cm) or CO ($\lambda$ = 2.6 mm). But this is the usual way of observing the structure from the Sun to the Galactic Centre, as well as the velocities associated with those sectors. As we have alread noted, interstellar extinction is always the biggest problem, so it is common to study other galaxies considered similar to Milky Way.

The \textit{bar} in the Milky Way is now fairly well established, owing to many studies that aimed in the same direction. Both observationally and theoretically -- i.e.\ by comparing observations with simulations -- the same conclusion is reached. Moreover, bars are a common feature of spiral galaxies. Even though the general rotation properties are very similar for SBb, SBc, Sb, and Sc galaxies, the barred ones are likely to produce more complex ripples in the observed rotation curves. However, if we really want to know their real influence on rotation curves, there are many parameters that need to be fixed (mass distribution, three axial lengths, etc). There is already much to study in this field. A good example is the length of the bar, which could be up to 7--8 kpc (L\'opez-Corredoira et al. \cite{CorredoiraCabreraGarzonHammersleyGonzalez_BAR2002}) with its consequently circular velocity distortion.

In spiral galaxies we can clearly see the well-known \textit{spiral arms} that give the name to this type of galaxy. These arms are also represented in our Galaxy. They are filaments that show a high amount of stellar formation with many O and B stars and consequently a large concentration of dust and gas in the interstellar medium.

All of these above-mentioned local substructures, together with interstellar gas and/or rings, may produce wavy ripples that cause the rotation curve to deviate from the usual exponential disc, especially in the optical disc.

Exploring the outer part of the disc is another difficult task, especially if we speak about low galactic latitudes. In these near-plane regions, there are available SDSS data that indicate some kind of stellar \textit{flare} (an increase in scale-height towards the outer Galaxy), even for long radial distances  around $\gtrsim$15 kpc, \cite{CorredoiraMolgo2014}. The flare is not another structure, just a continuous, smooth substructure of the disc. Of course the HI flare has also been explored in the last decade by many authors. A \textit{warp}, full of old stellar populations and gas, is also supported by observational studies. Star counts in the Galaxy hemispheres show clear asymmetries \cite{CorredoiraCabreraGarzonHammersley-WARP2002}.

\subsubsection*{The halo problem}

If we want to introduce the halo concept we need to speak about one of the most famous astronomers involved, Vera Rubin. In those times, where another areas was the centre of attention, Vera chose another field and fixed her attention on the stars of the Andromeda Galaxy. Since Newton's era, we have had a solid background for the kinematics of celestial objects. In our Solar System, the farther a body is from the Sun, the longer is the orbital period. The theory fitted perfectly with the observations, and they reproduced a clear descending curve. Somehow, the rotation of the Andromeda stars should have been reproducing this descending behaviour, but this was not what Vera found. On the contrary, the velocity kept almost constant along the radius, independently of the galactocentric distance. Why was it that what is true on the scale of the Solar system did not work on galactic scales? What should Rubin have done in that situation? Did the Newtonian equations and his theory need to be changed on galactic scales or, should be some kind of extra, extremely heavy or extremely abundant matter able to create enough gravity to make Newton's equations work? Vera opted for the second solution. She postulated that some kind of hidden mass or masses caused the the stars to rotate around the galactic centre confronting the tremendous centrifugal force produced by the high measured tangential velocities. And that is the initial idea of dark matter. There are anomalies in the behaviour and galactic dynamics, and if we want to understand them, we need to speculate about the existence of a kind of invisible matter whose origin and nature is already unknown. Vera performed many spectroscopic measures of HII regions and continued extracting conclusions (for instance \cite{Rubin1970,Rubin1982,Rubin1985}).

A paper published by Freeman in 1970, \cite{Freeman1970}, concerning the famous exponential disc is a great indicator of the need for extra mass in the disc. The rotation curve is not only determined just by the mass within the disc alone; it needs yet another contribution of some kind of invisible matter. In the following years, the observation of rotation curves led us to the concept of dark matter as the main gravitational ingredient of the halo.

But we cannot forget the luminous halo. This is mainly composed of `field stars', which represent the $\sim$0.98\% of the halo luminosity, leaving the rest to the globular clusters. Of course, the gravitational force produced by these objects is negligible compared with the full halo gravitational effect. The measured motion of the luminous halo, satellites, and even companion galaxies (Einasto \cite{Einasto1974}) indicates the dark halo envelopes the galaxy in all directions. Furthermore, the dark halo is usually considered to be close to a spherical shape (an ellipsoid in a more detailed description), and the main density equations depend only on the radius. Furthermore, calculating the circular velocity is very simple.

Most halo models have a reasonably good fit in the outer regions of the galaxy, levelling out to an almost flat rotation curve of $V_{\infty} \sim$ 200 km/s. But it is inside the optical disc radius where they usually change. The most well-known model is the isothermal one (Begeman, \cite{Begeman1985}), but also the NWF (Navarro-Withe-Frenk, \cite{NavarroWitheFrenk1996} ) and Burkert models (Burkert, \cite{Burkert1995}) are of great relevance in the literature. The main difference between them consists in the central cusp of the NWF model that produces the steepest rotation curve in the first kiloparsecs. However, the influence of the bulge is already clearly dominant at that distance, no matter what dark halo model we choose. Moreover, the halo is extended far away of the galactic disc; the flat curve produced holds until $\sim$30-50 kpc from the galactic centre.

Information can also be obtained by optical and infrared surface photometry. The photometric method uses the well-known mass-to-light parameter ($M/L$), that gives us an approximation of the mass distribution within the galaxy, and specifically about the disc. The surface-mass-density is measured and then, the $M/L$ ratio is applied. But in order to get this, we must already have an $M/L$ ratio. To get a good $M/L$ parameter we must also have, at the same time, an idea of the mass distribution beforehand; the mass distribution will change if we alter the $M/L$ ratio. And this parameter should be independent of the dark matter. Carignan \& Freeman in 1985 introduced the important concept of `maximum disc' by trying to manage without the unknown $M/L$ disc ratio, the disc mass contribution to the rotation curve is taken as large as possible to fit the observed rotation curve. That is, in order to derive the $M/L$ ratio we act as if we can fill the disc up of baryonic matter. In spite of this ambiguous parameter, the stellar component is a representative of the basic structure of the bulge and disc.

Despite the great amount of research and the great amount of detail discovered by the best surveys, the kinematics and their causes remain unknown, Or, at least, we do not understand them at all.

\section{Governing equations}

Here we are going to derive the main Newtonian equations that balance the centrifugal force versus the gravitational force. The reference frame origin will be at the 0-point (that is, the galaxy centre). All the lengths will be scaled to the outermost galactic radius $ R_g $, i.e.\ if we have the coordinate $X$ we can write $X = R_g x$. Since we will always work within the galactic radius, we have that $x \in [0,1]$.  Let us consider two points $Q(X,Y,Z)$ and $P(\hat{X},\hat{Y},\hat{Z})$ with the $Z$-axis perpendicular to the galaxy disc (defined by the plane given by the $XY$-axes). The distance between these two points will be given by

$$ d = |\mathbf{d}| = |\overline{QP}|=\sqrt{(\hat{X}-X)^2 + (\hat{Y}-Y)^2 + (\hat{Z}-Z)^2  } \spaceI .$$

But we are going to work with $r_S = d/R_g$,

\begin{equation}
r_S = \sqrt{(\hat{x}-x)^2 + (\hat{y}-y)^2 + (\hat{z}-z)^2  } \spaceI .
\end{equation}

Since we are only interested in comparing with the centrífugal force, we will just fix our attention in the $X$-coordinate (radial direction). If $\theta_x$ is the angle between the vector $\vec{d}$ and the X-axis the projection over the axis will be given by

$$ r_S \cos \theta_x = \hat{x}-x \spaceI . $$

Thus, by the Newton formula, the X-component of the force (in module) will be

$$ dF= G\dfrac{M_g dm}{d^2} = \dfrac{G M_g}{R_g^2} \dfrac{dm}{r_S^2}   \rightarrow  dF_X = dF \cos \theta_x  = \dfrac{G M_g}{R_g^2} \dfrac{\hat{x}-x}{r_S^3} dm \spaceI . $$

With $G = 6.6742 \times 10^{-11} m^3 Kg^{-1} s^2 $. In order to obtain the full force component over the point $P$ from a body $\mathcal{D} \subset \mathbb{R}^3 $ we will need to intagrate over $\mathcal{D}$. Since we know the mass differential can be split into the density function and the volume differential ($ dm = \hat{\rho} d\nu$) the integral will be

\begin{equation}
F_X = \dfrac{G M_g}{R_g^2} \int_{\mathcal{D}} \dfrac{\hat{x}-x}{r_S^3} dm = \dfrac{G M_g}{R_g^2} \int_{\mathcal{D}} \dfrac{(\hat{x}-x) }{r_S^3} \hat{\rho} d\nu \spaceI .
\end{equation}

Where $\hat{\rho} \equiv \hat{\rho}(\hat{x}, \hat{y} , \hat{z} )$ in terms of $M_g/R_g^3$ and $d\nu = d\hat{x} d\hat{y} d\hat{z}$. Since the set $\mathcal{D}$ will have the shape of a galaxy disc, cylindrical coordinates are appropriate. The height will be given by the $Z$-axis ($\hat{z} = \hat{h}$), and we note $\Delta h^2 = (\hat{h} - h)^2$ from now on. We will also consider the density distribution to be symmetric with respect to the $Z$-axis, that is $\hat{\rho} \equiv \hat{\rho}(\hat{r}, \hat{h} )$. Even though in the literature the capital letter $R$ is usually used for the radial cylindrical distance, here we will use $r$ (adopting {F\&G} notation for this) and we will leave capital letters for fixed parameters, constants, and integral limits.

\begin{equation}
\left. \begin{array}{l}
\hat{x} = \hat{r} \cos \hat{\theta} \\
\hat{y} = \hat{r} \sin \hat{\theta} \\
\hat{z} = \hat{h}
\end{array} \right\}
\entonces
\left\{ \begin{array}{l}
\abs{\pdiffI{(\hat{x}, \hat{y} , \hat{z})}{(\hat{r},\hat{\theta},\hat{h})}} = \hat{r}	\\	\\
\hat{x}-x = \hat{r}\cos \hat{\theta} - r \\ \\
r_S^2 = \hat{r}^2 + r^2 - 2\hat{r}r \cos \hat{\theta} + \Delta h^2 \spaceI .
\end{array} \right.
\end{equation}

Finally, the full Newtonian gravitational force produced by a disc through the $X$-axis is given by

\begin{equation}
F_X = \dfrac{G M_g}{R_g^2} \int_{-H/2}^{H/2} \int_{0}^{2\pi} \int_{0}^{1}    \dfrac{\hat{r}\cos \hat{\theta} - r}{ ( \hat{r}^2 + r^2 - 2\hat{r}r \cos \hat{\theta} + \Delta h^2)^{3/2}} \spaceI \hat{\rho}(\hat{r},\hat{h}) \hat{r} \spaceI d\hat{r} d\hat{\theta} d \hat{h} \spaceI .
\end{equation}

We have assumed the body to be a disc, where $H$ is the scaled thickness of the galactic disc. The dimensioned thick is $H_g = R_g H$. The scaled disc can be defined as

\begin{equation}\label{discSet}
\hat{\mathcal{D}} = \{ ( \hat{r},\hat{\theta},\hat{h} ) \in [0,1]\times[0,2\pi]\times [-H/2,H/2] \subset \mathbb{R}^3  \} \spaceI .
\end{equation}

On the other hand, we must have the centrifugal force at a distance $r$. Here, we will consider the centrifugal force just given by the radial distance $r$. As we have done with the radius, the circular velocity will be scaled by a characteristic velocity $V_0$ that usually is given by the maximum asymptotic rotational velocity. So we have $V = V_0 v$ and then

\begin{equation}\label{CentrifugalForce}
F_c = \dfrac{V^2}{ R }= \dfrac{V_0^2}{R_g} \dfrac{v(r,h)^2}{r} \spaceI .
\end{equation}

Thus, the equation that provides the force equilibrium must be $F_X + F_c = 0 $:

\begin{equation}\label{GoverningEquation3D}
\int_{-H/2}^{H/2} \int_{0}^{2\pi} \int_{0}^{1}  \dfrac{\hat{r}\cos \hat{\theta} - r}{ ( \hat{r}^2 + r^2 - 2\hat{r}r \cos \hat{\theta} + \Delta h^2)^{3/2}} \spaceI \hat{\rho}(\hat{r},\hat{h}) \hat{r} \spaceI d\hat{r} d\hat{\theta} d \hat{h} + A \dfrac{v(r,h)^2}{r} = 0 \spaceI ,
\end{equation}
where we get the relevant \textit{Galactic Rotation Number}.

\begin{equation}\label{A}
A = 	\dfrac{R_g V_0^2}{G M_g} \spaceI .
\end{equation}

This dimensionless parameter will have a continuous presence in all the sections. If we find a dimensionalized equation with the gravitational constant, the only thing we will have to do is substitute this identity and introduce the galactic rotation number.

Finally, we must remember the \textit{mass conservation equation} $ \int_{ \mathcal{D} } \rho d\mbox{Vol} = M $. If we normalize the equation, write it in cylindrical coordinates and apply the symmetric property of density distribution respect to the disc we get:

\begin{equation}\label{MCE}
4\pi \int_{0}^{H/2} \int_{0}^{1} \hat{\rho}(\hat{r},\hat{h}) \hat{r} \spaceI d\hat{r} d\hat{h}=1 \spaceI .
\end{equation}

\section{Target}

The idea of this master's dissertation is simple: taking this equation, discretizing and managing it to obtain results for rotation curves and/or density distributions. In the 
articles on which this dissertation is based, {F\&G} built a computational method based on the thin disc approximation and the equation \eqref{GoverningEquation3D} that we have just derived, but in its 2D form, as can be seen on the next chapter.

The first idea we are going to develop in chapter two will be the calculation of the rotation curves from a given density. This is something {F\&G} did not developed in their articles. We will do some tests and work through some examples, comparing them with other theoretical curves from disc distributions. We will encounter no trouble at all here.

What becomes problematic will be the `inversion problem'; that is, getting the density distribution given a theoretical or observed rotation curve. We find an integral equation to solve that requires a `kind matrix' to invert to avoid instabilities. {F\&G} are supposed to get a matrix that produces stable solutions whatever curve we use. They provide a method that is built in order to tackle a singularity produced in the nodes while the numerical integration is carried out. However, that method leaves us a stability problem that we are unable to solve completely. We even change the method, simplifying it. Throughout chapter three we tackle this obstacle by introducing certain modifications and we carefully explain the reason for the instability on various fronts.

After all this mathematical work, we go on to make some test and work some examples. We repeat some of the examples {F\&G} did to test our method. After that, we plot and comment on distributions obtained from different observed rotation curve data. Finally a simple dark disc deduction is made.

After all this work, the idea of extending these equations through the vertical axis inevitably occurred to us. Since the test and examples produced acceptable results for a thin disc, what output we will get if we introduce height into the equations? Certainly, the equation \eqref{GoverningEquation3D} is not fully used until the fourth chapter. First of all we perform an integral reduction using elliptic integrals to make the computation easier. We then need to build a two-dimensional grid to get a numerical representation of the height over the galactic plane. Including the property of the symmetric representation of the density distribution is another important point in reducing the cost of such a number of calculations within the code.

%==========================================================================================
% CHAPTER 2 ===============================================================================
%==========================================================================================

\chapter{Rotation curves from density distribution data}

\section{Numerical scheme}

Obtaining the circular velocity will not give us any problem; the scheme is simple, easy and fast to compute. Let's do the appropriate modifications to the original governing equations. If we take the integral contained in the original governing equations \eqref{GoverningEquation3D} and we do $\Delta h \rightarrow 0$ we will obtain
\begin{equation}
\int_{r=0}^1 \left[ \int_{\theta=0}^{2\pi} \dfrac{( \hat{r} \cos(\hat{\theta}) - r )}{( \hat{r}^2 + r^2 - 2 \hat{r} r \cos\hat{\theta} )^{3/2}} d \hat{\theta}  \right] H \rho(\hat{r}) \hat{r} d\hat{r} \spaceI .
\end{equation}

What we have now is a surface density $\Sigma(\hat{r}) \equiv H \rho(\hat{r})$, and the governing equations have just a bidimensional integral. Here we've got the F\&G problem:

\begin{equation}\label{GoverningEquation2D}
\int_{r=0}^1 \left[ \int_{\theta=0}^{2\pi} \dfrac{( \hat{r} \cos(\hat{\theta}) - r )}{( \hat{r}^2 + r^2 - 2 \hat{r} r \cos\hat{\theta} )^{3/2}} d \hat{\theta}  \right] \rho(\hat{r}) \hat{r} d\hat{r} + A \dfrac{v(r)^2}{H r} = 0 \spaceI .
\end{equation}

The reduction of the angle integral is required in order to get a faster computation. Using complete elliptic integrals, $K$ \& $E$ (for further description you can check appendix \ref{AppendixEllipticIntegrals}), we can reduce our equation into another which just contain one integral. This integral can be reduced using the identity

\begin{equation}\label{IntegralReduction2D}
\int_0^{2\pi} \dfrac{\hat{r}\cos \hat{\theta} - r}{ ( \hat{r}^2 + r^2 - 2\hat{r}r \cos \hat{\theta})^{3/2}} \spaceI d\hat{\theta} = \dfrac{2}{r} \left[ \dfrac{ E(k) }{ (\hat{r}-r)} - \dfrac{K(k)}{ (\hat{r}+r)} \right] \spaceI ,
\end{equation}
with an eccentricity given by

\begin{equation}\label{Eccentricity2D}
k^2=\dfrac{4\hat{r}r}{(\hat{r}+r)^2} = 1 - \dfrac{(\hat{r}-r)^2}{(\hat{r}+r)^2}; \spaceII k^2 \in [0,1] \spaceI .
\end{equation}

We will prove in the forth chapter a more general identity from which this one can be derived. Thus, the main equation \eqref{GoverningEquation2D} can be seen as

\begin{equation}
\int_{r=0}^1 \left[ \dfrac{E(k)}{\hat{r} - r} - \dfrac{K(k)}{\hat{r} + r} \right] \rho(\hat{r}) \hat{r} d\hat{r} + A \dfrac{v(r)^2}{2H} = 0 \spaceI .
\end{equation}

We will summarize here its basic numerical method implementation. First, we have to discretize the radius. Later a linear parameterization of radius and density will be done. Usually there's no need here of a huge nodal points to get a smooth curve. In case we have a very pronounced steep up following by another steep down, we could need a finest discretization. The set of nodes can be simply defined here as

\begin{equation}\label{Discretization2D}
\mathfrak{I} = \{  r_m \in [0,1] : m \in \{1,...,M\} \mbox{ , } r_{m+1}-r_m=s_m>0 \} \spaceI .
\end{equation}

And then, we have

\begin{equation}\label{GoverningEq2DDiscretized}
\sum\limits_{m=1}^{M-1} \int_{r_m}^{r_{m+1}} \left[ \dfrac{ E(k_{i}) }{\hat{r} - r_i} - \dfrac{K(k_{i})}{\hat{r} + r_i} \right] \rho(\hat{r}) \hat{r} d\hat{r} + \dfrac{v(r_i)^2}{2H} A = 0 \spaceI .
\end{equation}

For the radius we can make an isoparametric mapping and we build a linear one for the parametric density. The single parameter will be $\xi \in [0,1]$. The nodal points of the radius will be given by $r_m$ and the corresponding density nodes by $ \rho_m \equiv \rho(r_m) $.

\begin{equation}\label{Parameterization2D}
\left\{ \begin{array}{l}
\hat{r}_{m} \equiv \hat{r}_m(\xi)=(1-\xi)r_m + \xi r_{m+1} \\ \\
\diffI{\hat{r}_m}{\xi} = r_{m+1} - r_m = s_m
\end{array} \right.
\spaceII \mbox{\&} \spaceII
\hat{\rho}_{m} \equiv \hat{\rho}_{m}(\xi) = \rho_{m}(1-\xi) + \rho_{m+1}\xi \spaceI . \\ \\
\end{equation}

The parameterization of the density may not be fully necessary here but it will be helpful in later calculations and modifications. Thus, for every $r_i$ the numerical scheme of the equation \eqref{GoverningEquation2D} will look like

\begin{equation}\label{GoverningEq2DParam}
\sum\limits_{m=1}^{M-1} \int_{0}^{1} \left[ \dfrac{E(k_{i,m})}{\hat{r}_m - r_i} - \dfrac{ K(k_{i,n})}{\hat{r}_m + r_i} \right] \hat{\rho}_m \hat{r}_m s_m \spaceI d \xi + \dfrac{v(r_i)^2}{2H} A = 0 \spaceI .
\end{equation}

Furthermore we can easily calculate the velocity of every rotation curve in every radius node $r_i$.

\begin{equation}\label{Curves2DParam}
v(r_i) = \left( - \dfrac{2H}{A} \sum\limits_{m=1}^{M-1} \int_{0}^{1} \left[ \dfrac{E(k_{i,m})}{\hat{r}_m - r_i} - \dfrac{ K(k_{i,n})}{\hat{r}_m + r_i} \right] \hat{\rho}_m \hat{r}_m s_m \spaceI d \xi \right)^{1/2} \spaceI .
\end{equation}

The next step here is how do we calculate the integrals. The first problem that we face is the essential singularity located behind the second kind elliptic integral. We have selected the Legendre-Gauss quadrature because it only uses internal points with different weights (appendix \ref{AppendixGaussianQuadrature}). Thus, the value of $\hat{r}$ will never take the value $r_i$ at any operation. The number of nodal points we need to get a smooth curve, has not to be so big. With two hundred is enough for it. Perhaps if the initial steep is so pronounced a larger number could be required. We usually have used five hundred nodal points in the next plots.

From now on, if we want to compare the obtained rotation curves with other ones coming from another density distribution or another theoretical formula, the selection of this number will be relevant. If we have not a Galactic Rotation Number given by the distribution itself (like the Mestel and Freeman-Mestel ones) we have to calculate it according to some fixed values. The main parameter we are going to choose here will be the radius $R_g$. For instance, if we consider the full disc mass ($\sim 6.5 \times 10^{10}$ M$_{\Sun}$) and we have $R_g$ in $m/s$, $A$ can be calculated by
 
$$A_d = \dfrac{V_0^2 R_g}{G M_{d}} = 5.6092 \times 10^{-21} R_g \spaceI .$$

\section{Some density distribution models}

The aim of these section is to check if this basic scheme to get the rotation curves is enough accurate to take the method into account. We will try with some basic theoretical cases to check if our results fit with the well known curves from the huge amount of literature about this topic. Since we need to test our numerical model and its code we are going to test some relevant density profiles.

\subsection{Freeman exponential disc}\label{SubsectionFreemanExpDisc}

The exponential density distribution of a disc was proposed for Freeman \cite{Freeman1970} in 1970 through a classic study of the photometry. After applying the corresponding mass-to-light ratio the shape of this exponential disc would be simply given by:

\begin{equation} \label{ExponentialDiscFreeman}
\rho(r) = \rho_0 e^{-r/R_d} .
\end{equation}

In our nondimensionalized variables, the total integrated mass will be equal to 1 for all the distributions at a radius $r=1$. In case of an exponential disc it can be easily calculated:

\begin{equation}\label{ExponentialDiscMass}
M_{\mathcal{D}}(r) = \int_{ \mathcal{D}(r) } \rho \spaceI d\mbox{Vol} = 2\pi H \rho_0 R_d \left[ R_d - e^{-r/R_d}(r+R_d) \right] \spaceI ,
\end{equation} 
where $\mathcal{D}(r)$ is the disc of radius $\hat{r} \in [0,r]$ as we have described at \eqref{discSet}. We can get the value of the central density of the disc, $\rho_{d,0}$, just by selecting the scale radius of the disc $R_d$ whose value has been taken from \ref{Sofue2013} as 3.5 kpc. The computed rotation curve for this distribution can be seen in figure \ref{FigureExponentialCurves}. In order to test it, we have compared it with other curves. The first one is the obligatory one, the theoretical curve known for an exponential disc. We have taken the one deduced by \textit{Binney \& Tremaine} \cite{Binney} using Modified Bessel Functions.

\begin{equation}\label{ExponentialDiscVelocityBessel}
v_c^2 (r) = 4 \pi G \Sigma_{d,0} R_d y^2 [  I_0(y)K_0(y) - I_1(y)K_1(y)  ] \spaceII \mbox{where} \spaceII y=\dfrac{r}{2R_d} \spaceI .
\end{equation}

$I_{\nu}(z)$ and $K_{\nu}(z)$ are the Modified Bessel Functions of first and second kind. The surface density can be express as $\Sigma_{d,0} = H \rho_{d,0}$ and we have to take care of substituting the GRN of equation \eqref{A} to eliminate dimensions from the equations. We have use here the value given by Sofue \cite{Sofue2013} for the Mass of the full disc, $M_d=6.5 \times 10^{10}$.

What we can extract from figure \ref{FigureExponentialCurves} is that the computed rotation curve fits perfectly with the theoretical one making impossible to distinguish between them. Therefore the thin disc approximation that we are testing works in this relevant example.

\subsection{Spherical distributions and the projected mass distributions}

Another rotation curve we have selected to plot is the one given for a spherical distribution. Getting the rotation curves here comes directly from the Gauss theorem due to what we have got is a spherical distribution. Remember that $r_S$ denotes the spherical distance (not the cylindrical). The rotation curves will be given by:

\begin{equation}\label{VelocitySphere}
v_c^2 (r_S) = \dfrac{G M(r_S)}{r_S} \spaceI .
\end{equation}

In this case, the central density changes its value. We can calculate it by the same way, through the integrated mass of an exponential sphere making $r_S=1$ and determining the bulge size.

\begin{equation}
M_S(r_S) = 4\pi\rho_0 R_d \left[  2R_d^2 - e^{-r_S/R_d}( r_S^2 + 2 r_S R_d + 2 R_d^2 )  \right] .
\end{equation}

A spherical distribution will show a lower rotation velocity at figure \ref{FigureExponentialCurves}, although for large distances both of them will tend to the keplerian rotation velocity as we can check in \textit{Binney \& Tremaine}, \cite{Binney}.

\begin{figure}[h!]
\centering
\includegraphics[width=0.8\textwidth]{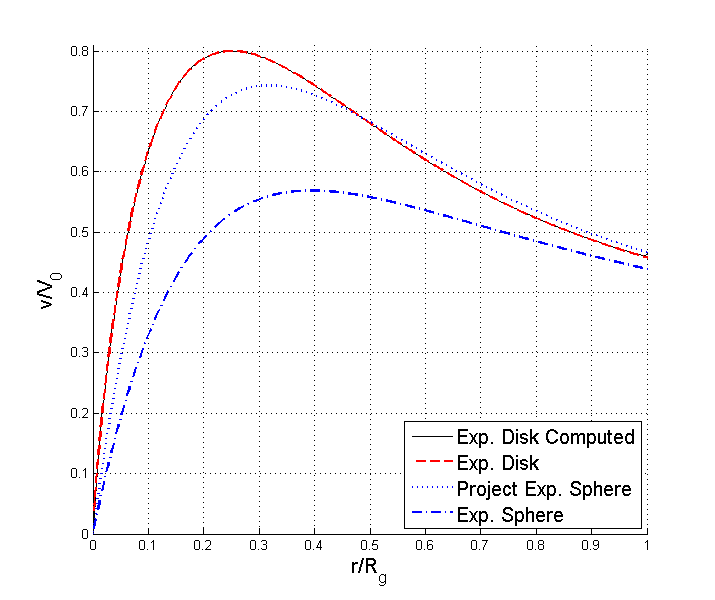}
\caption{Rotation curves of exponential distributions. All the distribution have been adjusted to have the same total integrated mass, $R_g=30$ kpc and $R_d = 3.5$ kpc. The lines of the analytical solution of an exponential distribution (dashed line) and the computed one are superimposed. The other two ones show the comparison between the velocity produced by an exponential sphere (pointed line) and the computed curve for its density projection within a thin disc (points).}
\label{FigureExponentialCurves}
\end{figure}

The profile that we obtain with this exponential distribution does not look like the observed ones in most of the galaxies; the velocity does not keep approximately constant at large distances. On the other hand, these velocities fit perfectly with the systems where the dark matter models are not involved (like the globular and open clusters).\\

The spherical distribution will lead us to another question that will be useful to illustrate the next and last section about getting the rotation curves from density. If we compress the mass of the spherical distribution inside a disc and apply the method, what are we getting? The projected mass density can be easily calculated by

\begin{equation}\label{ProjectedDistribution}
\Sigma(r) = \int_{\mathbb{R}^+} \rho(r,h) dh \spaceI .
\end{equation}

We have calculated it numerically. Remember $r$ is the cylindrical radius again here. Note that we must divide this surface mass density by $H$ to ``distribute'' the mass through the disc. That is $\rho(r) = \Sigma(r)/H$.

Clearly, we will not get  the same rotation curve. As we can check in the figure \ref{FigureExponentialCurves} the profile will be greater. As the other curves, for enough large radius the curve will also approach to the keplerian profile, but in this case the projected mass density will stand out the other ones at the outer radius.

\subsection{Mestel \& Freeman-Mestel distributions} \label{SubsectionFMcurves}

When we observe many rotation curves of a spiral galaxy we can easily abstract a common pattern in the shape of them. Departing from the center of the galaxy, the curve will rise up with a strong slope in order to stabilize around a certain value of the radius keeping the velocity almost constant. It is easy to find in the literature an idealized description of the observed rotation curves:

\begin{equation}\label{MestelCurve}
v(r)=1-e^{-r/R_c} \spaceI ,
\end{equation}
where $v$ has again no dimension itself; its value is relative to a maximum orbital velocity $V_0$. If we consider the limit $R \rightarrow 0$ there is a solution for the density within a disc. This solution is known as Mestel's disc (obtained by Mestel \cite{Mestel1963}). This distribution is expressed in a simpler way for F\&G:

\begin{equation}\label{MestelDisc}
\begin{array}{lcl}
\rho(r) & = & \dfrac{A}{2 \pi H r} \left[ 1 - \dfrac{2}{\pi} \sin^{-1}(r) \right]  \\ \\
A & = & \dfrac{1}{\int_0^1 \left[ 1 - \frac{2}{\pi} \sin^{-1}(\hat{r}) \right] d\hat{r}} = \dfrac{\pi}{2} = 1.5707 \spaceI .
\end{array}
\end{equation}

We can see there is a singularity at $r=0$. Since the exponential disc formula from \textit{Freeman} \cite{Freeman1970} has not that problem and, at the same time, describes reasonably well the central section of the galaxy that is dominated by bulge matter, F\&G has the idea of mixing both formulae. The exponential shape of equation \eqref{ExponentialDiscFreeman} will give us the correct shape until some certain value and, from there, the Mestel distribution will describe the disc. They called it the \textit{Freeman-Mestel} distribution.

\begin{equation}\label{FreemanMestelDistribution}
\begin{array}{l}
% ------------------------------------------------------------------ Density
\rho(r) =
\left\{ \begin{array}{lcl}
\rho_0 e^{-r/R_d} & \mbox{if} & r \in [0,\tilde{R}_c) \\
\dfrac{A}{2 \pi H r} \left[ 1 - \dfrac{2}{\pi} \sin^{-1}(r) \right]  & \mbox{if} & r \in [\tilde{R}_c,1]
\end{array}\right.
\\ \\ % ------------------------------------------------------------------ R_b value
R_d = \left\{  \dfrac{1}{\tilde{R}_c} + \dfrac{2}{  \pi [ 1 - 2 \sin^{-1}(\tilde{R}_c)/\pi ]  \sqrt{1-\tilde{R}_c^2}  }   \right\}^{-1}
\\ \\ % ------------------------------------------------------------------ rho_0
\rho_0 = \dfrac{A}{2 \pi H \tilde{R}_c e^{-\tilde{R}_c / R_b} } \left[ 1 - \dfrac{2}{\pi} \sin^{-1}(\tilde{R}_c) \right]
\\ \\ % ------------------------------------------------------------------ A
A = \left[ 2\pi H  \sum\limits_{m=1}^{M-1} \int_0^1 \rho_{A=1}(\xi) \hat{r}(\xi) \dfrac{d\hat{r}}{d\xi} d\xi  \right]^{-1}
\end{array}
\end{equation}
\\
Freeman-Mestel is the basic density distribution that F\&G use to test their method \cite{FengGallo2011}. Furthermore it is important for us to replicate a similar test to check if our model works as it should be. The galactic rotation number $A$ is given by the mass conservation equation \eqref{MCE}. The parameter $\tilde{R}_c$ becomes relevant here and, specially, in the next chapter. It looks like to have a similar physical meaning of $R_c$ in equation \eqref{FreemanMestelDistribution}. Somehow they represent the galactic core. Despite this, there is no a real mathematical link between them. We can check in figures \ref{FigureCurvesMestel} how these differences take form in the rotation curves if we take the same value for both parameters. Moreover, the galactic rotation number obtained from Freeman-Mestel model is a little bit greater compared with the one that we will compute in the next chapter if we do $R_c=\tilde{R}_c$. Anyway, the point is we can easily watch how the Freeman-Mestel computed rotation curves are getting closer to the theoretical ones as we get close to zero ($R_c,\tilde{R}_c \rightarrow 0$), and we can see it perfectly in figure \ref{FigureCurvesMestel}, where one curve almost overlaps the other for $R_c = \tilde{R}_c = 0.001$. In addition, the galactic rotation number has the same behaviour, and $A_{FM} \rightarrow A_{Mestel} = 1.5707$

\begin{figure}[h!]
\centering
\includegraphics[width=0.5\textwidth]{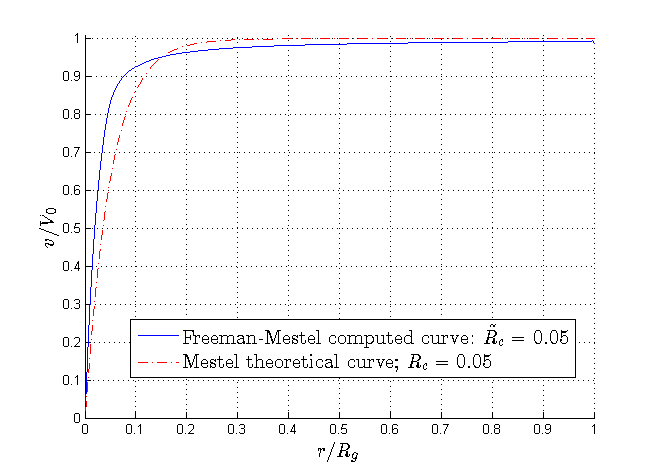}\includegraphics[width=0.5\textwidth]{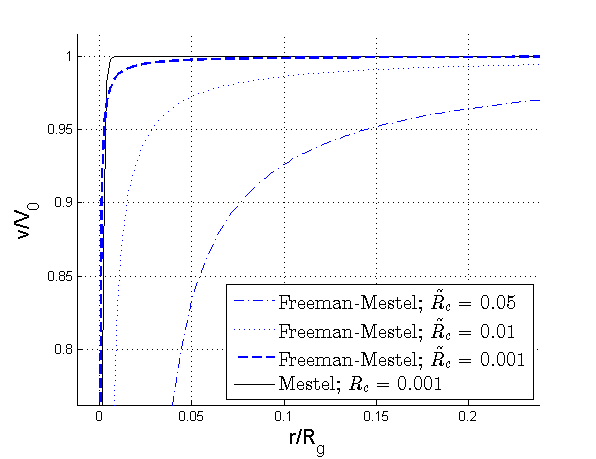}
\caption{Mestel and Freeman-Mestel computed curves. \textit{Left:} The blue line is the computed curve from the Freeman-Mestel density distribution of eq. \eqref{FreemanMestelDistribution} with $\tilde{R}_c = 0.05$ and the red dashed line is the idealized curve of eq. \eqref{MestelCurve} with $R_c=0.05$. \textit{Right:} Some computed curves from Freeman-Mestel profile with $\tilde{R}_c$ values tending to zero. At the limit, there is an idealized curve with $R_c=0.05$ }
\label{FigureCurvesMestel}
\end{figure}

As we can observe, the values from the galactic rotation number are similar to the ones that we have calculated for the total galactic mass. If we fix our attention in our method \eqref{Curves2DParam}, we are supposed to select an appropriate galactic rotation number. In contrast, this distributions give us a predetermined one. That implies these distributions let the rotation velocity keep almost constant along most of the galactic radius just allocating all the galactic mass within the radius. That is, the disc does not need any additional mass to describe reasonably well the galactic rotation. That is something that F\&G \cite{FengGalloDarkMatter} discuss widely in other interesting papers and we will have time later to comment about it.

\subsection{Flared Disc}\label{SubsectionFlaredDisc}

Despite of what we have commented in the last section, here we are interested in how the flares affect the rotation curves in the last section of the visible disc. As we have checked the basic shape of the curves will not be highly modified, so it could be interesting to plot some comparisons here of flared or non-flared exponential disc.

First of all we must introduce the flared disc proposed by \textit{López-Corredoira \& Molgó} \cite{CorredoiraMolgo2014}. The disc was calculated with the SDSS-SEGUE data. Although they have different kinds of data including the Metallicity gradient along the galactic radius, we better take the simplest one: We select all the data. There is also a distinction between thin and thick disc, that we will not take into account: The mass/stars ratio from the thin disc to the thick disc is represented by $f_{thick}/f_{thin}=0.09$. Density is expressed in cylindrical coordinates. We can summarize it here as:

\begin{equation}\label{FlaredDisc}
\rho(r,h) = \rho_{\Sun} \mathcal{A}(r) \spaceI \exp{\left(  - \dfrac{\abs{h}}{\mathcal{B}(r)} + \mathcal{C}_R(r) + \mathcal{C}_h(r)  \right) } \spaceI .
\end{equation}

The amplitude $\mathcal{A}(r)$ makes the density decrease in the plane zones, spreading the mass all over the vertical scale. The component $\mathcal{C}_R(r)$ defines the exponential falling off with the radius. The last component, $\mathcal{C}_{h}(r)$, provides the disc models with a ``hole'' in the inner disc that is controlled by the parameter $h_{r,hole} \approx 3.74$ kpc.

\begin{equation}\label{FlaredDiscComponents}
\mathcal{A}(r) = \dfrac{h_{z}(R_{\Sun})}{h_{z}(r)}; \spaceII \mathcal{C}_{R}(r) = \dfrac{R_{\Sun}-r}{h_{r}}; \spaceII \mathcal{C}_{h}(r) = h_{r,hole} \left( \dfrac{1}{R_{\Sun}} - \dfrac{1}{r} \right). 
\end{equation}

Finally, the component relative to the flare will depend of the height scale function of the thin/thick disc. We are noting $R_{ft} \approx 6.9  \spaceI$ kpc as the radius where the thick disc flare starts. In case we want a non-flared disc, we would just have to do $k_{i}=0$. 

\begin{equation}\label{FlaredDiscScales}
\begin{array}{lcl}
\mathcal{B}_r = h_{z}(r) & = &
\left\{ \begin{array}{lcl}
h_{z}(R_{\Sun}) & \mbox{if} & r \in [0,R_{\Sun}) \\ 
h_{z}(R_{\Sun}) \left( 1 + \sum\limits_{i=1}^{2} k_{i} (r-R_{\Sun})^i \right) & \mbox{if} & r \in [R_{\Sun},\infty) \\
\end{array} \right.
\end{array}
\end{equation}

Even though we have not plotted it, the output of the program with $h_{hole}\approx3.74$ kpc is clearly remarks how the model works. Since what we do is projecting the mass, there is a high peak of gravitational mass around 3-4 kpc in contrast with the inner 2 kpc, where the mass amount is almost zero. This fact provokes in the program there is no rotation velocity until 2-3 kpc. That makes sense due to what we have is a similar situation to the one provided for a ring of mass. Perhaps the intuition makes us think the gravitational force would make the particle inside the ring move towards the center or (remembering the Gauss Theorem) just keep it in the same place, with all the forces cancelling each other. On the contrary, the particle will move away from the center, making impossible to make the gravitational force cancel the centrifugal force, which is the background of the method.

First of all we can clearly observe the model depends on height definition values. We will be able to treat this more carefully in the forth chapter. But we can make an interesting approximation here. Since we need just the radial distance to apply the method, we are going to use the projected mass distribution of the model. If we use \eqref{ProjectedDistribution} with our flared density distribution, we can easily find an analytical solution:

\begin{equation}\label{FlareProjected}
\Sigma(r) = 2 \rho_{\Sun} \mathcal{A}(r)\mathcal{B}(r) \exp{(\mathcal{C}_R(r) + \mathcal{C}_h(r))} = 2 \rho_{\Sun} h_{z}(R_{\Sun}) \exp{(\mathcal{C}_R(r) + \mathcal{C}_h(r))}.
\end{equation}

This can be applied not only to the thin Disc, also to the thick disc in case it has a similar mathematical representation. As we can easily check there is no dependence of $h_{z}(r)$, so if we compare a flared and non-flared disc we will obtain the same velocity profile. This distribution just spread the mass of the disc along the Z-axis departing from a certain radius. However, this can give us a light hint of how the flare can softly affect the rotation curve in spite of the lack of stellar matter in the inner disc.

%===========================================================================================
% CHAPTER 3 ================================================================================
%===========================================================================================

\chapter{Density distribution from rotation curve data}

\section{Numerical scheme}

Here we have the inverse problem. Our goal here is to get the surface density distribution through all the radius just by knowing the rotation curve data. The journey to get it is a little bit more complicated than the one that we have already done in the last chapter.

From the parameterized governing equation \eqref{GoverningEq2DParam} for every $i \in \{1,..,M\}$ we can see we have $M$ equations, one for every node $r_i$. In order to set the equations up we should fix our attention in the parameterization of the density in equations \eqref{Parameterization2D}, the density through an interval ($[r_m,r_{m+1}]$) is split in two, where every part is multiplied by one of the two considered nodes $\rho_m$ that are considered constants for the integral and they can be taken away. Those nodes are going to be precisely our unknown variables:

\begin{equation}
\int_{0}^{1} \left[ \dfrac{E(k_{i,m})}{\hat{r}_m - r_i} - \dfrac{ K(k_{i,n})}{\hat{r}_m + r_i} \right] \hat{\rho}_m \hat{r}_m s_m \spaceI d \xi = \rho_{m} \Gamma_{(i,m)}^{(1-\xi)} + \rho_{m+1} \Gamma_{(i,m)}^{(\xi)} \spaceI ,
\end{equation}
where the coefficients of the matrix system will be composed of integrals of the form
 
\begin{equation}\label{GammaSubElements}
\Gamma_{(i,m)}^{(\xi)}  =  \int_0^1 \left[ \dfrac{E(k_{i,m})}{\hat{r}_{m} -r_i} - \dfrac{K(k_{i,m})}{\hat{r}_{m} + r_i} \right] \xi \hat{r}_{m} s_{m} \spaceI d\xi \spaceI .
\end{equation}

What leave us with $M$ equations and $M$ unknown variables. But we may want to consider the galactic rotation number as another unknown variable, so that we would need another equation. And for that is why we need the Mass Conservation equation \eqref{MCE}. We have to discretize and parameterize it (see equations \eqref{Parameterization2D} and \eqref{Discretization2D} ); after both transformations we obtain

\begin{equation}\label{MCEParam}
\sum\limits_{m=1}^{M-1} \int_0^1 \hat{\rho}_m \hat{r}_m s_m d\xi = \dfrac{1}{2\pi H} \spaceI .
\end{equation}

The corresponding coefficients related with the Mass Conservation Equation have the same structure:

\begin{equation}
\int_0^1 \hat{\rho}_m \hat{r}_m s_m d\xi  = \rho_m \Delta_{(i,m)}^{(1-\xi)}  + \rho_{m+1} \Delta_{(i,m)}^{(\xi)}  \spaceI .
\end{equation}

This integrals can be also easily solved without numerical integration (what we have is a simple polynomial integral):

\begin{equation}\label{DeltaElements}
\begin{array}{lclcl}
\Delta_{(i,m)}^{(\xi)} & = & \int_0^1 \xi \hat{r}_{m} s_{m} \spaceI d\xi & = & s_m \left( \dfrac{1}{2} r_m + \dfrac{1}{3} s_m  \right) \\ \\
\Delta_{(i,m)}^{(1-\xi)} & = & \int_0^1 (1-\xi) \hat{r}_{m} s_{m} \spaceI d\xi & = & s_m \left( \dfrac{1}{2} r_m + \dfrac{1}{6} s_m \right) \spaceI .
\end{array}
\end{equation}

After adding this equation to the set of governing equations and organizing the different terms, what we find is an integral equation; a linear system $(M+1)\times (M+1)$ of the form $\Gamma \vec{\rho}_A = \gamma$. Where the components are $\Gamma \in \mathcal{M}_{M+1 \times M+1}(\mathbb{R})$ and $\gamma, \vec{\rho}_A \in \mathcal{M}_{M+1}(\mathbb{R}) $. The unknown vector is $\rho_A = (\rho_1,...,\rho_M,A)$ and the independent vector is $\gamma=(0,...,0,1/2\pi H)$. The system sketch can be visualized here as

\begin{equation}
\left( \begin{array}{ccc|c}
 & & &  \\
 & & &  \\
 & \Gamma_{M}  & & \dfrac{v(\vec{r})^2}{2H} \\
 & & &  \\
 & & &  \\ \hline
 & \mbox{Mass conservation equation} & & 0
\end{array} \right)
\left( \begin{array}{c}
\\
\\
\vec{\rho} \\
\\
\\ \hline
A 
\end{array}  \right)
=
\left( \begin{array}{c}
\\
\\
0 \\
\\
\\ \hline
1/2\pi H 
\end{array}  \right)
\end{equation}

We summarize the matrix elements here, which are easily described by integrals \eqref{GammaElements} and \eqref{DeltaElements}. What this elements precisely describe is the kernel function of the integral equation.

\begin{equation}\label{GammaElements}
\begin{array}{l}
% Gamma elements -------------------------------------------------------------------
\mbox{If  } i\in \{1,...,M\} \spaceII \left\{ \begin{array}{lcl}
m=1 & \rightarrow & \Gamma_{(i,1)} = \Gamma_{(i,1)}^{(1-\xi)} \\ \\
m\in \{2,...,M-1\} & \rightarrow & \Gamma_{(i,m)} = \Gamma_{(i,m-1)}^{(\xi)} + \Gamma_{(i,m)}^{(1-\xi)}  \\ \\
m=M & \rightarrow & \Gamma_{(i,M)} = \Gamma_{(i,M)}^{(\xi)} \\ \\
m=M+1 & \rightarrow & \Gamma_{(i,M+1)} = \dfrac{v(r_i)^2}{2H}
\end{array} \right.
% Delta elements ---------------------------------------------------------------------
\\ \\
\mbox{If  } i=M+1 \spaceII \left\{ \begin{array}{lcl}
m=1 & \rightarrow & \Gamma_{(M+1,1)} = \Delta_{(i,1)}^{(1-\xi)} \\ \\
m\in \{2,...,M-1\} & \rightarrow & \Gamma_{(M+1,m)} = \Delta_{(M+1,m-1)}^{(\xi)} + \Delta_{(M+1,m)}^{(1-\xi)} \\ \\
m=M & \rightarrow & \Gamma_{(M+1,M)} = \Delta_{(M+1,M)}^{(\xi)} \\ \\
m=M+1 & \rightarrow & \Gamma_{(M+1,M+1)} = 0
\end{array} \right.
\end{array}
\end{equation}
\\

Now the system is fully described. This is the real mathematical core of F\&G \cite{FengGallo2011,FengGallo2014} articles. Solving this system should provide us of the density distribution at the nodes.

\section{The stability problem}\label{SectionStabilityProblem}

\subsection{Numerical representation of the kernel function}

But if we want to solve this system and get an useful solution, the system must be stable. We can observe how matrix size depends of the number of nodes, so what we will usually get is a huge one. The truth is that we have serious problems here, as we will check later, so let us comment the different obstacles we have found in the $\Gamma$ matrix.

The first problem we found is that the first row of the matrix will be zero due to we have $r_1=0$. This fact will make the eccentricity equation \eqref{Eccentricity2D} equal to zero. Because of this, the values of the complete elliptic integrals will be the same, $E(0) = K(0) \approx 1.5708$. That is, the values of the components of the matrix $\Gamma$ elements \eqref{GammaSubElements} will be cancelled ($\Gamma_{1,m}^{\xi} = \Gamma_{1,m}^{1-\xi} = 0$). To avoid this problem and get rid of this first equation of the system, we simply have supposed $\rho_1 = \rho_2$. Then, we only have to sum the first column to the second one and eliminate the first column and the first row to have now a $M \times M$ system. The elimination of the first row (equation) is something not indicated in the original F\&G articles \cite{FengGallo2011,FengGallo2014} but it looks to be a clear required step.

Once we have solved this issue, we just need to solve the matrix, and the explanation requires an essential break about how we can calculate the matrix elements, which are the principal characters here. These elements are given by integrals that has to be numerically calculated.

\begin{enumerate}
\item As we have indicated in the last chapter we decided to apply directly the 6-point Gauss quadrature to the integrals \eqref{GammaSubElements} that compose the $\Gamma$ elements. This method avoid the nodal point so the essential singularity is supposed to be properly avoided.

\item In F\&G articles \cite{FengGallo2011,FengGallo2014} are a few hints indicated for a method development to transforms these integrals in order of dodge the singularities. The integrals are not calculated directly. They use approximations that even involve two-dimensional integrals to calculate the elements where the singularity appears (where we used again $6 \times 6$ Gauss quadrature method for 2D integrals as we described in appendix \ref{AppendixGaussianQuadrature} ). The development of those equations is quite long and is not included in this paper. However we have studied, developed, written carefully and finally built a code to use it. In fact, that is the first code we wrote.
\end{enumerate}

\subsection{System matrix structure}

In case we are using a typical density distribution of a full galaxy, the solution must have very different values, from very close to zero values (outermost radius) to a number that is many orders of magnitude over it (bulge section). This fact will be shown on the coefficients of the equations (the matrix elements). The matrix that we obtain has a general structure that we can describe as follows. The main diagonal (we note $d_0$) has huge values compared to the the upper triangular matrix elements ($d_m$ with $m>0$). In fact, the upper triangular matrix has positive values. The diagonal $d_{-1}$ has very close to zero values and the elements of diagonals $d_m$ with $m < -1 $ have negative values. But the problem is in diagonal $d_{-2}$, which has absolute values very similar to the diagonal $d_{0}$. The absolute values behind $d_{-2}$ are negligible. Somehow this structure an these values has clear physical meaning, where the negative values represents the attraction of the inner/outer masses that creates the equilibrium on a specific radius node. The real problem here is that we get the same structure independently of what element calculation method we use from the ones described in the last subsection.

The point is that we obtained an unstable solution using both ways. And it does not matter what we choose, the main problem is the same: We are suppose to get a diagonally dominant matrix and we do not get it. F\&G affirms they obtain this kind of matrix. A diagonally dominant matrix is defined by

\begin{equation}
\abs{a_{ii}} = \sum\limits_{i \neq j } \abs{a_{ij}}
\end{equation}

This property guarantees us a bounded condition number \cite{HillarShaoweiWibisono}, which is a measure of the system stability.

\subsection{Solving the linear system}

The next step is solving the linear system. Getting the inverse matrix usually is not the best idea. Even more if the nodes set is huge and the matrix is not precisely stable. So we must check in the literature the many different ways of solving linear systems. There are more stable solvers and more unstable ones. The iterative solving methods usually requires sparse matrix or, at least, the diagonally dominant matrix property to work properly. What we have then are the direct methods. We have computed the most relevant ones. After all, we finally decided to use the MATLAB/OCTAVE standard solve which is based of ``LU'' method after considering many options due to we have the more accurate results together with the total pivoting Gaussian method. Besides, F\&G comment they refer the main matrix $\Gamma$ as a ``Jacobian'' matrix with a bounded condition number (check appendix \ref{AppendixConditionNumber} again) for the Newton-Raphson method. This method is destined for solving nonlinear systems, where the Jacobian matrix have real meaning. You can find a complete description of the most famous direct and iterative techniques in \cite{Infante} and how the main system solver function works (including its flowchart) at \cite{MATLABmldivide}.

\subsection{Main cause and location of the instability in the data input}\label{SubsectionMainCause}

We can also observe that the instability looks to appear harder in some specific cases. In the next section we are going to plot some examples and we are going to treat them individually, but after all these computing results and different examples, it is easy to find a relation with the initial velocity steep in all the cases. The more pronounced the harder instability.

In addition to this, the instability specially affects the last section of the radius in relative terms. If the initial steep is enough pronounced we do not obtain a clear smooth curve there. We use here a simple and orthodox method to solve this problem: The mean of two points of the solution. It's just enough to do it once to get a perfect smooth curve in all the cases. However, we have done it twice in order to fix the original nodal points. In spite of we get this smooth curve, the problem is not clearly solved: The final solution we obtain usually show an early falling off when the steep is really pronounced.

We will treat every case individually, taking the result from the Mestel idealized curves in the next section as a guide examples for different initial steep profiles.

\section{Freeman-Mestel distribution}

As we have already done in the last chapter, we take as a starting point Freeman-Mestel distributions. In the last chapter we presented and plotted a set of idealized rotation curves \eqref{MestelCurve} with its analytical solution of the density at $R_c \rightarrow 0$ given by equation \eqref{MestelDisc}. The first characteristic that fix our attention in the Mestel Disc equation is the lack of the galactic core parameter $R_c$ because we have already taken its limit. At the same time there is a clear singularity at $r=0$. That is the singularity that the Freeman-Mestel distribution equation \eqref{FreemanMestelDistribution} eliminates.

\begin{figure}[h!] 
\centering
\includegraphics[width=0.7\textwidth]{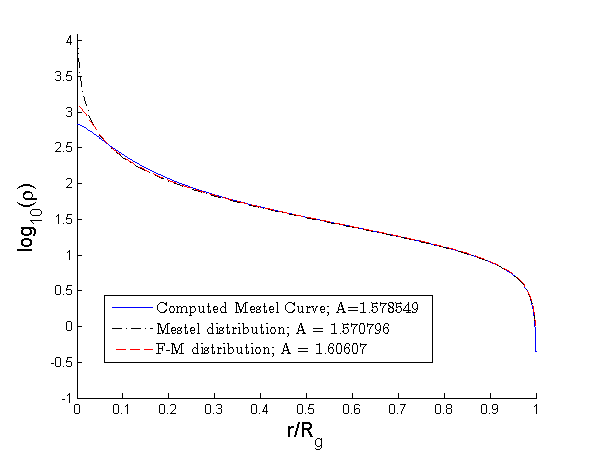}
\caption{Density distributions. The black dashed-pointed line describe the Mestel distribution. The red dashed line describe the Freeman-Mestel distribution with $\tilde{R}_c = 0.05$. And the blue line is the computed density distribution of an idealized rotation curve given by equation \eqref{MestelCurve} with $R_c=0.05$.}
\label{FigureF-MDensities}
\end{figure}

It is easy to imagine that the first figure we must plot is the comparison between the Mestel and the Freeman-Mestel distribution together with the computed density distribution of an idealized rotation curve. We can watch this in figure (\ref{FigureF-MDensities}). The first thing we can comment here is about the clear density disparity at the bulge region. The disparity of the Mestel disc has a clear motive, given its singularity. The other two distributions depend on two parameters that we have already spoken about, $R_c$ and $\tilde{R}_c$ (in both cases, we assign the same value, $0.05$). As we said, they have not real mathematical relationship between them, so their influence is not the same on the density plot. On the other hand the region far from the galactic center is very similar and the lines overlap one to each other.

Another point is connected with the general shape of all the plotted densities. This shape corresponds to the standard spiral galaxies density profiles. Speaking in $\log_{10}$ terms, there is a highly steep descending slope in the bulge region, an almost constant distribution at the medium part (with a logical weak descending density) until the very end of the outermost radius, where the density has a precipitous fall to 0.

In addition to all the mentioned points, we cannot forget the program also calculates the GRN. Since we are working with dimensionless magnitudes, if we select a galactic radius, and a characteristic velocity we can easily calculate the total galactic mass. For instance, if we computed $A = 1.5788$ and we consider a radius of $R_g = 16$ kpc and a characteristic velocity of $V_0=220$ km/s we get $M_g \approx 1.4 \times 10^{11}$ M$_{\Sun}$.

Everything in figure \ref{FigureF-MDensities} looks fine, but here we have the computed density distribution from the idealized curve with $R_c=0.05$. But, what happens if we reduce the value of $R_c$ in the input velocities? The initial steep becomes more and more vertical so, as we have indicated in the last section, the instability increases towards the outer radius. How can we measure some kind of bound where the instability give us critical results? Somehow, the Freeman-Mestel distribution should tend to the Mestel distribution as $\tilde{R}_c \rightarrow 0$. Furthermore the Galactic Rotation Number should do the same and it should tend to the original value given by Mestel: $A=\pi/2=1.5708$.			

\begin{figure}[h!] 
\centering
\includegraphics[width=0.8\textwidth]{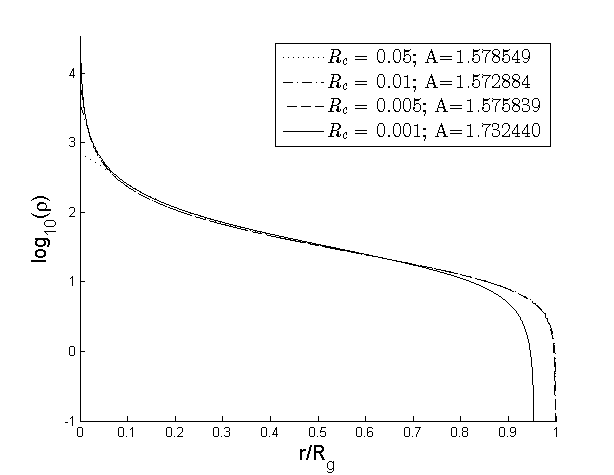}
\caption{Density distributions computed from idealized curves of eq. \eqref{MestelCurve} with different core scale $R_c$. The stability becomes a real problem somewhere between $R_c=0.005$ and $R_c=0.001$, where the solution is a disaster.}
\label{FigureStabilityProblem}
\end{figure}
		
\begin{table}[h!]
\centering
	\begin{tabular}{|c|c|c|c|c|} \hline
$R_c$ & Feng \& Gallo & Our computed values \\	\hline
0.05  & 1.5777 & 1.578549  \\ \hline
0.01  & 1.5719 & 1.572884  \\ \hline
0.005 & -	   & 1.575839  \\ \hline
0.001 & -      & 1.732440  \\ \hline
	\end{tabular}
	\caption{Computed Galactic Rotation Number (see equation \eqref{A}) from Mestel idealized curves of equation \eqref{MestelCurve} with different $R_c$ values. \textit{Left}: Computed by F\&G code. \textit{Right}: Computed by our code.}
	\label{TableStabilityProblem}
\end{table}

We can see in table \ref{TableStabilityProblem} that the value of A increases when we compute a distribution from an idealized curve with $R_c =0.005$. This is a good hint about there is something wrong, and the value for $R_c = 0.001$ confirms an unacceptable instability. Moreover, if we plot the curves we can observe what this instability produces. Taking into account the modifications we have indicated in subsection \ref{SubsectionMainCause}, we obtain a smooth curve and since the values of the density tends to zero very fast, if we plot the density without applying the $log_{10}$, the differences are negligible. But if we apply the logarithm, the contrast becomes disastrous at the outermost disc at $R_c=0.001$. The perturbations of the instability forces the mean of the values to cross the X-axis, becoming negative ones. In logarithm terms we interpret we have 0 density. That is, for critical values the lack of stability makes the density distribution plunge to zero too quickly. On the contrary, the case defined by $R_c=0.005$ has a very acceptable plot where the last density collapse occurs, making the discrepancy with the other lines imperceptible as you can see in figure \ref{FigureStabilityProblem}.

Another interesting test would be going backwards on some computed rotation curve from Freeman-Mestel density. Avoiding dangerous values of $\hat{R}_c$ for stability, the obtained density form the computed rotation curve overlaps the original Freeman-Mestel distribution, so we have not found any complication here.

\section{Milky Way density distribution}\label{SectionMilkyWayDensityDistribution}

If we want to test our method we should go backwards on our computed results. If we do it with the curve obtained from Freeman exponential disc or the analytical curve solution, basically they overlap each other except at the last section ($r \gtrsim 0.7 R_g $), where the distribution obtained for our program force the density to gradually fall off to zero due to the action of a self-gravitating disc (figure \ref{FigureExpDiscBackwards} ) . As we have indicated before, instability only strongly emerges where there is a critical vertical initial velocity. Somehow, this test, is a confirmation of it.

\begin{figure}[h!] 
\centering
\includegraphics[width=0.7\textwidth]{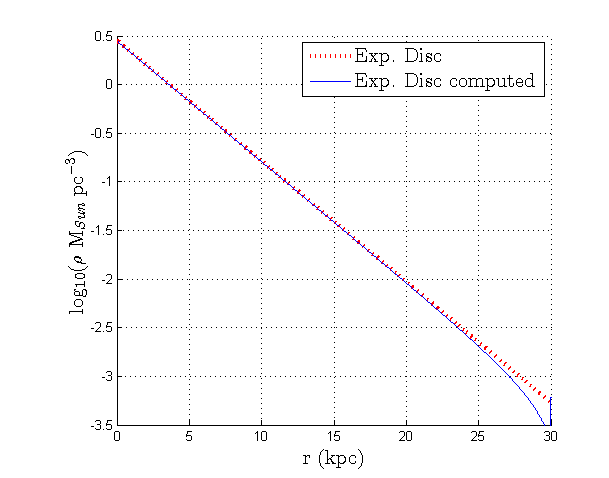}
\caption{Exponential disc original distribution density (red pointed line) and exponential disc distribution computed from the corresponding rotation curve velocity (blue line).}
\label{FigureExpDiscBackwards}
\end{figure}

But now that we have seen how the method reacts to a theoretical rotation curve shape and the well known exponential disc, we can try to use it with some realistic observational data of the Milky Way. We have taken as main source the Sofue data web page \cite{SofueCurves2013}. The original data surveys comes from 2009, \cite{SofueCurves2009}. The velocities have many different sources: HI tangent velocity method, CO tangent, HI tangent, CO and HII regions and HI thickness (all the authors and dates of this surveys are indicated in \cite{SofueCurves2009}). All those sets of data points have been smoothed by Gaussian-running mean what gives the solution we can observe in figure \ref{FigureMWdistribution} (up-left). We can be tempted to introduce this curve directly in the program, but with a initial first view we can assume there will be problems. Certainly, as we have commented in subsection \ref{SubsectionMainCause} the initial steep could bring us instability here. And we cannot solve it directly. Milky Way is not precisely the best galaxy in this sense. The initial steep is very pronounced if we compare it with other ones like NGC 3198 or NGC 2708 (you can see both examples in \cite{FengGallo2011}). But we are mainly interested in the outer part of the disc and since the density distribution obtained beyond 0.3- 0.4$R_g$ is almost the same for different initial steep (figures \ref{FigureF-MDensities} and \ref{FigureStabilityProblem}), we decided to modify the original bulge shape till certain radius to avoid the instability problem without compromising seriously our main scientific target. Still, very steeped curves will causes instability so we must take care of our bulge domain.

The idea is very simple. We build an artificial bulge within some given radius $R_{bulge}$. Beyond the bulge radius we will simple make an interpolation of the Sofue curve. Spline method have been considered, but the last section of the curve is so flat that the spline does not give us a good approximation. On the contrary, the bulge is built by a simple spline with three points of support. The initial point is zero valued. The intermediate point will be located at $R_{bulge}/2$ and its value will correspond to the maximum of the bulge in the original data. The last point will obviously take the real value on $R_{bulge}$. It is recommended to take one more point to make the transition smoother on $R_{bulge}$ when we calculate the spline, even though we do not use it later. You can see how we work the bulge problem out in figure \ref{FigureMWdistribution} (upper plots). On this situation, the shorter we take $R_{bulge}$, the less we modify the observed distribution, but we always have the danger of instability. We have computed the distribution with $R_{bulge}=1.5$ kpc observing no instability like the one produced in figure \ref{FigureStabilityProblem}. However, from $R_{bulge}=2$ kpc to $R_{bulge}=1.5$ kpc we observed small increase of the Galactic Rotation Number (order around $10^{-3}$) so we stopped decreasing the parameter $R_{bulge}$. In spite of all the last  arguments, the order is already quite small to indicate any real stability problem.

We also wanted to test our program with other more wiggled data from 2012 Sofue source \cite{SofueCurves2012}. In this case we have fitted a polynomial through least-squares. The variation of data radius steps brought us some difficulties: The data discretization beyond 10 kpc is very thick and variable, so the usual least-squares fit does not work properly beyond r$\sim$20 kpc. Just by selecting an optimal polynomial degree and outermost radius we avoided any fit problem. Later we apply the same bulge modification of the last paragraph to the fitted curve. The result can be seen at figure \ref{FigureMWdistribution} (up). At this point we must admit we did not take into account the enormous uncertainties of the data. On the other hand, we find here a perfect test for our method.

Before we begin with the description of the output, it is a good idea to check how the GRN values react to the total mass data. The GRN obtained for every data set are $A_{2013}=2.17$ (Sofue 2013 data \cite{SofueCurves2013} till $R_g=24$ kpc) and $A_{2012}=1.85$ (Sofue 2012 data \cite{SofueCurves2012} till $R_g=20$ kpc). If we consider $V_0=220$ km/s we can calculate the total galactic mass:

\begin{equation}
\begin{array}{lcl}
M_{2013}(R_g = 24 \mbox{ kpc}) & = & \dfrac{V_0^2 \cdot 24}{G \cdot A_{2013}}= 1.24 \times 10^{11} \spaceI \mbox{M}_{\Sun}  \\ \\
M_{2012}(R_g = 20 \mbox{ kpc}) & = & \dfrac{V_0^2 \cdot 20}{G \cdot A_{2012}}= 1.22 \times 10^{11} \spaceI \mbox{M}_{\Sun}.
\end{array}
\end{equation}

The values obtained are in concordance with F\&G results \cite{FengGallo2011}. They obtain for shorter radius ($r\approx15.3 $ kpc) a mass of $M_g = 1.10 \times 10^{11}$ M$_{\Sun}$, what is in very good agreement with the star count of Sparke \& Gallagher, \cite{SparkeGallagher2007}. Our interest is a little far beyond that radius, and our values should be greater. The total galactic mass within a 20 kpc sphere is supposed to be $1.5 \times 10^{11}$ M$_{\Sun}$ (Sofue at \cite{Sofue2013}) and if we reach 50 kpc we have $2.9 \times 10^{11}$ M$_{\Sun}$ (Gibbons, 2014). These values let us give dimensions to the dimensionless volume density we have been using until now. We can make an approximation of it just by multiplying by $M_g/R_g^3$. For instance, the solar neighbourhood ($R_{\Sun} \approx 8$ kpc) has $\rho_{\Sun}\approx 0.4$ M$_{\Sun}$ pc$^{-3}$. That is a very accurate value, compared with usual approximations of 0.2-0.3  M$_{\Sun}$ pc$^{-3}$ (bulge+disc+halo) or 0.18-0.3  M$_{\Sun}$ pc$^{-3}$ (bulge+disc)  (values from \cite{Sofue2013}). We should not forget this method implies that the density of the thin disc of 0.01 $R_g \approx 240$ pc must support the whole curve. Besides, the SMD gives a value ($\Sigma_{\Sun} = \rho_{\Sun} \times 240 \approx 1.06 \times 10^2$  M$_{\Sun}$ pc$^{-2}$) between the bulge+disc value ($0.89 \times 10^2$  M$_{\Sun}$ pc$^{-2}$ ) and the bulge+disc+halo value ($4.2 \times 10^2$  M$_{\Sun}$ pc$^{-2}$). The values for the 2012 distribution are also very similar ($\rho_{\Sun}\approx 0.4$ M$_{\Sun}$ pc$^{-3}$ and $\Sigma_{\Sun} \approx 1.23 \times 10^2$  M$_{\Sun}$ pc$^{-2}$).

\begin{figure}[h!]
\centering
\includegraphics[width=1\textwidth]{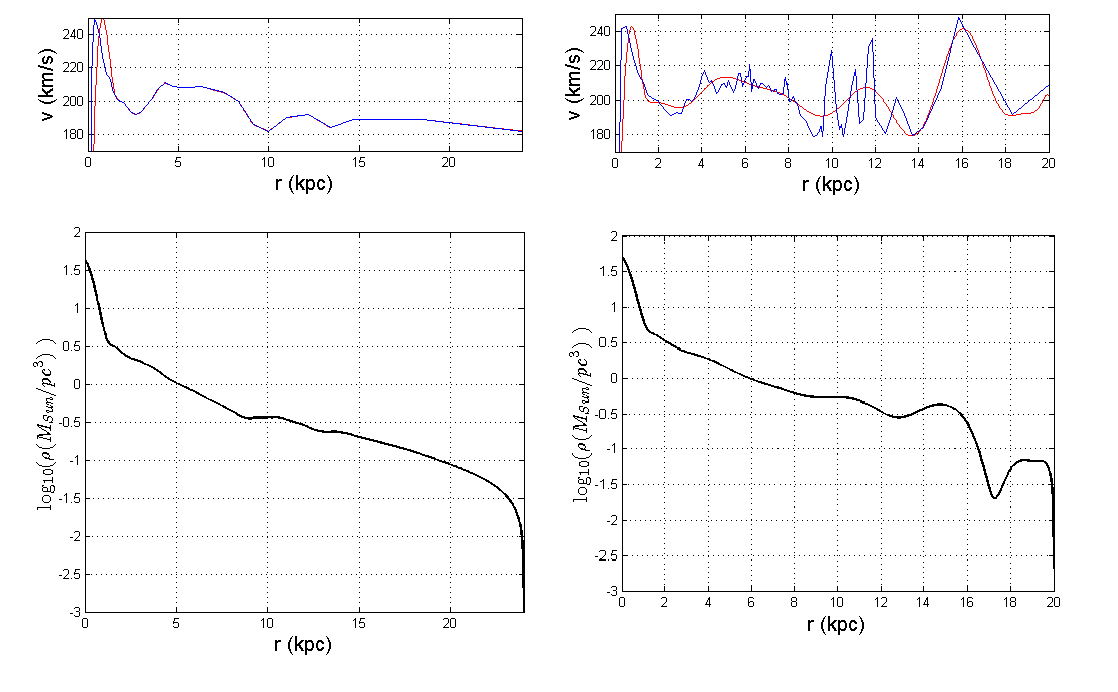}
\caption{Rotation curves with modified bulges and computed density distribution for two data set (2013 data \cite{SofueCurves2013} on the \textit{left} and 2012 data \cite{SofueCurves2012} on the \textit{right}). \textit{Up}: Original data (blue line) and modified bulge and fitted data (red line). \textit{Down}: Computed log$_{10}$ density distribution in M$_{\Sun}$ pc$^{-3}$.}
\label{FigureMWdistribution}
\end{figure}

Finally the density distribution shape that we have obtained is shown in figure \ref{FigureMWdistribution}, together with both rotation curves (bulge-modify and original one). The modified bulge seems not to affect significantly the behaviour of remote distance from the center. The wiggles and ripples produced in the density happen to meet precisely in the ripples of the rotation curve. Where the velocity increases over an idealized flat curve, the density data obtained corresponds with an upper rip in a similar section like in the section just after the bulge ($r\sim2.5$ kpc) or just after $r\sim 10$ kpc. At last, there is another small velocity rise just around $r\sim 13-14$ kpc and then, the curve descend gradually till the given radius is done, when the density fall off to zero. The conception that we have of the galaxy has a clear fit with the curve and its mass distribution.

If we consider the 2012 data, beyond 10 kpc from the galactic center, we know the dispersion of the velocities measured in our galaxy is quite considerable as we can check precisely in \cite{SofueCurves2012}. In addition, when we plot these data, there is a clear difference from the already treated rotation curve given in \cite{SofueCurves2013}. The rotation curve starts a clear ascent around r$\sim$15 kpc that peaks around r$\sim$15-16 kpc. What we get here leave us another conclusion about how the method works. In this last case, we can observe how the density distribution is spread by this method in a self-gravity thin disc that holds such as rotation curve. The rotation curve upper rips output density upper rips just before the velocity ones (around 1-2 kpc before). Perhaps it was merely observable in 2013 data, but here, specially in the outer disc, becomes clear. In 2013 data, the pure flat curve given beyond 15 kpc does not let us observe any wiggle there. A discussion about the perfection of the linear system is taken into account. The product of the instability is precisely an early fall of in the density distribution, so we must take care of this result. However, we have considered this handicap when we have modified the bulge.

\section{Galaxy decomposition}

In order to prepare the explanation of the next section, it is convenient to describe the model we are going to use on subsequent modifications. The model we are going to use is the classical \textit{de Vaucouleurs bulge + exponential disc}. The description of our bulge just by de Vaucouleurs law would be clearly inadequate in a deeper study of it. A more specific bulge would be composed of the Central Black Hole and two distinct sphere exponential laws provoked by a galactic spherical core and the spheroidal bulge (Sofue, 2013 \cite{SofueCurves2013}). We can give here a brief description of the spherical bulge distribution. de Vaucoulers bulge (1958) comes from a surface brightness profile law. From that profile a Surface Mass Density profile (SMD) has been deduced:

\begin{equation}\label{VaucouleursBulgeSigma}
\Sigma_{b} (r) = \Sigma_{be} \exp \left[ -\kappa \left(  \left( \dfrac{r}{R_{b}} \right)^{1/4} - 1 \right) \right] \spaceI .
\end{equation}

The Half-Mass scale radius takes the value $R_{b}=0.5$ kpc. The corresponding SMD at this $R_b$ is given by $\Sigma_{be} = 3.2 \times 10^2$ M$_{\Sun}$pc$^{-2}$. In order to calculate the spherical velocity given by equation \eqref{VelocitySphere} through the projected mass distribution of equation \eqref{ProjectedDistribution} we need to calculate the mass enclosed within a sphere of radius $r$. Given the SMD, the volume mass density is calculated by the following integral (Sofue, \cite{Sofue2013}):

\begin{equation}
\rho_b(r) = \frac{1}{\pi} \int_R^{\infty} \diffI{\Sigma_b(x)}{x} \dfrac{1}{\sqrt{x^2-r^2}} \spaceI dx = \frac{\kappa}{4 R_b \pi} \int_R^{\infty} \left( \dfrac{x}{R_{b}} \right)^{-3/4} \dfrac{- \Sigma_b(x)}{\sqrt{x^2-r^2}}  \spaceI dx \spaceI .
\end{equation}

The exponential disc will be calculated as we have indicated in subsection \ref{SubsectionFreemanExpDisc}, departing from equation \eqref{ExponentialDiscVelocityBessel}. In this case we used directly the central SMD of the disc, $\Sigma_{d,0} = H \rho_{d,0} = 8.44 \times 10^2$ M$_{\Sun}$ pc$^{-2}$.

Since we want to get a complete model of the galaxy we have also included the isothermal dark halo (Begeman,  \cite{Begeman1985}). This dark halo is the best fitted to this galaxy decomposition selection. Finite density mass in the center is establish here, like the Burkert model (\cite{Burkert1995}) does, which corresponds better with the observational data. Other dark halos like the NFW model (\textit{Navarro-Frenk-White}, \cite{NavarroWitheFrenk1996}) are derived from N-Body simulations and represent the set of ``cusp models'' (that have cusp of dark matter in the galactic core) which is not considered here. The isothermal spherical density distribution can be expressed as

\begin{equation}\label{HaloIsothermal}
\rho_{iso} (r_s) = \dfrac{ \rho_{iso}^0 }{ 1 + (r_s/R_h)^2} \spaceI ,
\end{equation}
and its velocity will be given by

\begin{equation}
v_{h}(r) = v_{\infty} \sqrt{ 1 - \left( \frac{R_h}{r} \right) \tan^{-1}\left( \frac{r}{R_h} \right)  } \space I .
\end{equation}

The value for the halo radius is taken from \cite{SofueCurves2013}, $R_h \approx 12$ kpc. And the parameter $v_{\infty}$ determines the flat rotation at infinity. Its value is related with the other parameters: $v_{\infty} = \sqrt{4 \pi G \rho_{iso}^0 Rh^2} \approx 200$ km/s. Finally, all the velocity components have the vectorial relation

\begin{equation}\label{VelocityComponents}
v_{total}^2 = v_{b}^2 + v_{d}^2 + v_{h}^2 \spaceI .
\end{equation}

You can watch the decomposition of theoretical galactic curve in figure \ref{FigureMWcomponents}, together with the original curve data from Sofue \ref{SofueCurves2013}. We have added some indicative data like the velocity that the \textit{bulge + exp. disc} model needs to reach the observational one.

\begin{figure}[h!]
\centering
\includegraphics[width=0.5\textwidth]{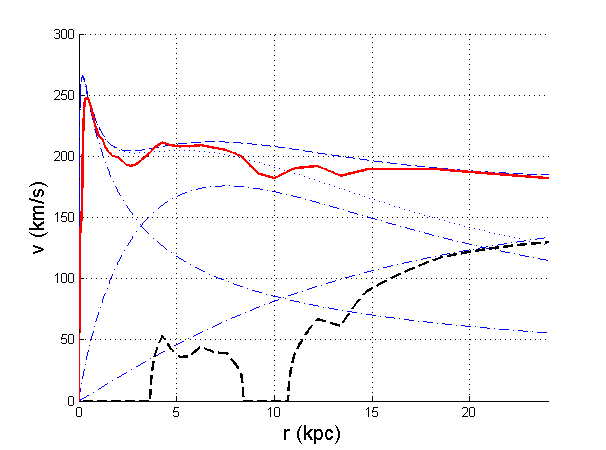}\includegraphics[width=0.5\textwidth]{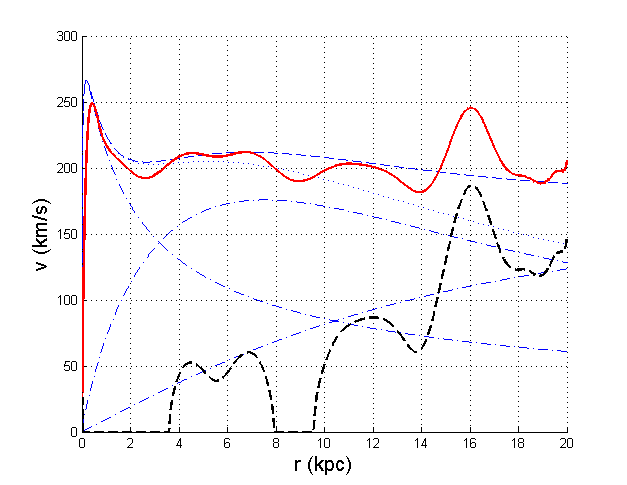}
\caption{Rotation curve decomposition. \textit{Left:} 2013 data \cite{SofueCurves2013}. \textit{Right} 2012 data \cite{SofueCurves2012}. The thick red line is the fitted original observed data, blue dash-pointed lines are the decomposition \textit{bulg + disc + halo} and blue thin dashed line is the sum of different component for all of them as a theoretical full galaxy rotation curve. Finally, the black thick dashed line represents the subtraction of the \textit{bulge+disc} from the original observed data}
\label{FigureMWcomponents}
\end{figure}

You can watch the decomposition of theoretical galactic curve in figure \ref{FigureMWcomponents}, together with the original curve data from Sofue \ref{SofueCurves2013}. From this figure we can extract some comments. When we subtract the exponential disc and the bulge velocities to the observed ones (using equation \eqref{VelocityComponents} ), the first property that get our attention is that we obtain two components. Both of them has several rips throughout its shape. The main reason for this is that we have not taken into account the wiggles of the galaxy disc. Prominent upper and lower dips produced by arms and other substructures are not considered in the equation but the can be observed in the rotation curve data usually considered. Even other perturbations can be treated by considering ``ring waves'' superposed on the original exponential disc. The equations and possible models are exposed in \cite{SofueCurves2009}. Moreover it could be the halo itself has substructures that have effect on the disc dynamics (Hayashi and Chiba \cite{Hayashi2006} ; Bekki and Chiba \cite{Bekki2006}). However this studies suggest the substructures would affect the outer rotation curve. Perhaps the irregularities from 2012 Sofue data \cite{SofueCurves2012} around 15-16 kpc has something to do with it, but that is something far beyond of our target here.

In this specific case where bulge has been subtracted, it takes all the density with it, representing the full rotation curve through this first section. As we have commented, since we are using de Vaucouleurs bulge which does not the best fit to our galaxy, this result can be expected.

The first component we have found (3 kpc $\lesssim r \lesssim$ 8 kpc) leaves no clear difference between the to data sets. We can distinguish at least two peaks at the 2013 data and just two peaks for the 2012 data due to we are using fitted one. The relation to the dark halo must be discussed here. In general, it looks like the velocity produced by the halo is some kind of maximal velocity to this first component. And that is exactly what it must be expected. After that, there is another ``zero section''. This is clearly given by the instability of this galaxy model decomposition. Perhaps an exponential disc that falls earlier or some lower dip implementation could help to this representation. Finally, in the 2013 data, the second component appear (r $\gtrsim$ 10 kpc) into the shape of a curve that grows seeking the halo curve that tends to stabilize at $v_{\infty}\approx 200$ km/s at the outer part of the disc. The case for the 2012 data has the same behaviour, a low dip around 15 kpc. But beyond that radius, we find a peak that clearly stand the halo velocity curve out. Due to what we have commented just a few lines above this is what we expected.

\section{Dark disc density distribution}\label{SectionDarkDisc}

Since we are working on F\&G articles, it could be interesting work on their ideas. In F\&G \cite{FengGalloDarkMatter} they speak about the density interstellar medium (ISM) as great supporter of the rotation curves in the disc. Some of their ideas come from the decreasing rate of star formation beyond a certain radius where the lack of dense molecular clouds becomes decisive. On the other hand, outside the optical disc, gas clouds (both ionic, atomic and molecular form) and dust \textit{``are likely to provide enough mass for explaining the observed rotation curve. Only a few (baryonic) atoms per cm$^3$ in terms of the average number density could be sufficient''}. That is, they suggest the ISM could still hold enough mass in longer radius to keep the rotation curves like we observe them.

The first test that came to our mind is the following: We take the density distribution outputted from the last rotation curves exposed in the last section \ref{SectionMilkyWayDensityDistribution}, and we subtract the densities produced by a theoretical bulge and exponential disc. What we will obtain is the distribution of the lack/excess of density that we need to support the observed rotation curve. That is what you can see (in absolute value) in figure \ref{FigureMWdarkdiscrho}. The apparent excess of mass behind that radius, as we have commented, could be produced by the decomposition method and the ripples of the disc which have not been implemented. However, in both data sets, the lack of density becomes relevant beyond $r\sim 11$ kpc. That is exactly what should be expected. Until the end of the considered radius, the density difference has a maximum around $r\sim 15-16$ kpc of $\Delta \rho \lesssim 0.1$ M$_{\Sun}$ pc$^{-3}$. The descending curve from there is produced just because we consider a self-gravitational disc.

We also have plotted another line in the lower plots (red dashed line) of figure \ref{FigureMWdarkdiscrho}, that correspond just to an illustrative exercise. We have taken the curve produced by the isothermal halo and we have apply the code to it. What gives us is an indicator of how much disc density is needed to hold the flat rotation curve.

\begin{figure}[h!]
\centering
\includegraphics[width=0.5\textwidth]{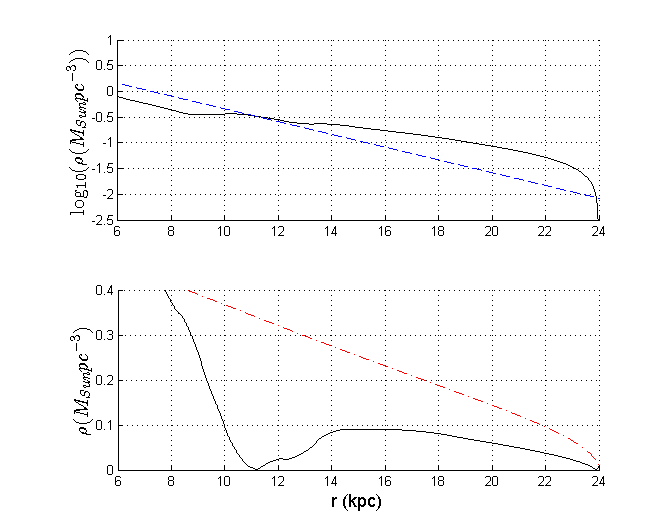}\includegraphics[width=0.5\textwidth]{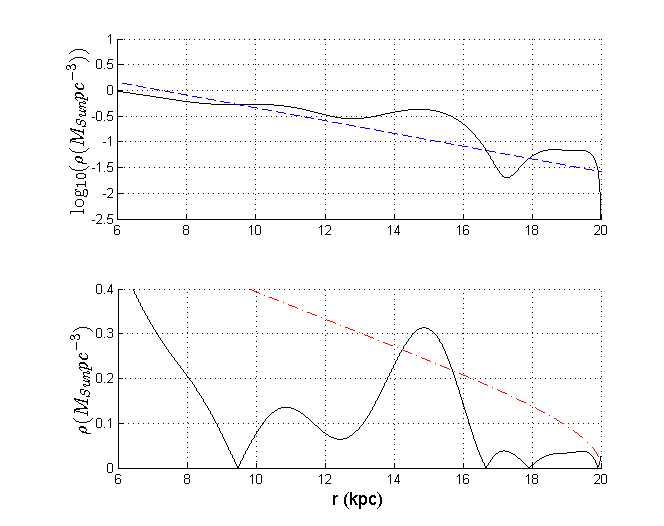}
\caption{Spatial density distribution difference from obtained Milky Way data and classical \textit{bulge + exp. disc} for two data set (2013 data \cite{SofueCurves2013} on the \textit{left} and 2012 data \cite{SofueCurves2012} on the \textit{right}). \textit{Up}: Milky Way density distribution obtained (black line) and \textit{bulge + exp. disc} model densities (blue dashed line).\textit{Down}: Absolute value of the difference between the above distributions in M$_{\Sun}$ pc$^{-3}$.}
\label{FigureMWdarkdiscrho}
\end{figure}

%===========================================================================================================================
% CHAPTER 4 ================================================================================================================
%===========================================================================================================================

\chapter{Method expansion to 3D}\label{ChapterMethodExpansionTo3D}

After all the previous work, it is a natural question how can we expand this method through vertical axis. Even more if we have contemplated the governing equation \eqref{GoverningEquation3D} with a height component. Of course that was made on purpose, having this last chapter in mind. However a few considerations you maybe have realized should be made here.
The first is clear: If we fix our attention on the centrifugal force equation \eqref{CentrifugalForce}, we can see we just have used the cylindrical radius instead the spherical one. As far as we consider a point over the galactic plane the force in the vertical axes must take its place.

\section{Integral reduction by elliptic integrals}

In order to get an easier computation, our new goal is to eliminate the $\theta$-dependency in the integral using elliptic integrals properties. The integral we must treat is

\begin{equation}\label{Integral3D}
\mathtt{I}(r,\hat{r},\Delta h^2) = \int_0^{2\pi} \dfrac{\hat{r}\cos \hat{\theta} - r}{ ( \hat{r}^2 + r^2 - 2\hat{r}r \cos \hat{\theta} + \Delta h^2)^{3/2}} \spaceI d\hat{\theta} = \int_0^{2\pi} \dfrac{\hat{r}\cos \hat{\theta} - r}{ | \mathbf{\hat{r}-r}|^3 } \spaceI d\hat{\theta} \spaceI .
\end{equation}

To get a more general view, let us skip all the calculations by now, write the final solution and do some observations:

\begin{equation}\label{IntegralReduction3D}
\mathtt{I}(r,\hat{r},\Delta h^2) = \dfrac{2}{r} \left[ \dfrac{ (\hat{r}+r)(\hat{r}-r) + \Delta h^2 }{ [ (\hat{r}-r)^2 + \Delta h^2]   \sqrt{(\hat{r}+r)^2 + \Delta h^2}}  E(k)  - \dfrac{1}{ \sqrt{(\hat{r}+r)^2 + \Delta h^2}} K(k)   \right] \spaceI .
\end{equation}

\begin{description}
\item[obs 1: ] $K(k)$ and $E(k)$ are the first and second kind \textit{complete elliptic integrals} defined in appendix \ref{AppendixEllipticIntegrals}. The \textit{elliptic integral modulus} or \textit{eccentricity} is given here by

\begin{equation}
k^2=\dfrac{4\hat{r}r}{(\hat{r}+r)^2 + \Delta h^2} = 1 - \dfrac{(\hat{r}-r)^2 + \Delta h^2}{(\hat{r}+r)^2 + \Delta h^2}; \spaceII k^2 \in [0,1] \spaceI .
\end{equation}

\item[obs 2: ] Now, in our governing equation, we have just to solve a 2-dimensional integral.

\begin{equation}
\int_{-H/2}^{H/2} \int_{0}^{1} \mathtt{I}(r,\hat{r},\Delta h^2)  \spaceI \hat{\rho}(\hat{r},\hat{h}) \hat{r} \spaceI d\hat{r} d\hat{\theta} d \hat{h} + A \dfrac{v(r,h)^2}{r} = 0 \spaceI .
\end{equation}

\item[obs 3: ] If we make ($\Delta h^2 \rightarrow 0$), this will lead us to the well known identity \eqref{IntegralReduction2D} for the bidimensional case, that we have already used for the pure thin disc problem.

\item[obs 4: ] Numerical computations have been made in order to check the integral identity, finding no difference between the values obtained through numerical integration and the formula with elliptic integrals.

\end{description}

\section{Rotation curves calculus}

\subsection{Numerical Scheme in 3D}

Now we face the problem of calculating this numerically. In the scope of this discretization there is a double numerical integral, so we need to build a 2D-grid. Going through the entire radius we will have a system of $(M+1) \times (2N +1)$ equations of the form:

\begin{equation}\label{GoverningEquationsNoSymmetry}
\sum\limits_{m=0}^{M-1} \sum\limits_{n=-N}^{N-1}  \int_{r_m}^{r_{m+1}} \int_{h_n}^{h_{n+1}} \mathcal{S}^{(i,j)}(\hat{r} , \hat{h}) \hat{\rho}(\hat{r},\hat{h})\hat{r}  \spaceI d\hat{r} d\hat{h} + \frac{A}{2} v(r_i,h_j)^2 = 0 \spaceI ,
\end{equation}
where

\begin{equation}\label{Scaligraphic}
\mathcal{S}^{(i,j)}(\hat{r} , \hat{h}) = \left[ S_E^{(i,j)} E(k^{(i,j)}) - S_K^{(i,j)} K(k^{(i,j)}) \right](\hat{r} , \hat{h}) \spaceI ,
\end{equation} 

\begin{eqnarray}
S_K \equiv S_K(r,\hat{r},\Delta h^2) & = &  \dfrac{1}{ \sqrt{(\hat{r}+r)^2 + \Delta h^2}} \spaceI ,  \\
S_E \equiv S_E(r,\hat{r},\Delta h^2) & = &  \dfrac{ (\hat{r}+r)(\hat{r}-r) + \Delta h^2 }{ [ (\hat{r}-r)^2 + \Delta h^2]   \sqrt{(\hat{r}+r)^2 + \Delta h^2}} \spaceI .
\end{eqnarray}

This situation will leave us with a huge amount of equations, but we will always take here symmetrical density distributions regarding the galactic plane. If we use the density symmetric hypothesis we can reduce these equation (so the unknowns) to the half. It is obvious that, if we have this symmetry, then the radial velocities will be the same at both sides of the disc (at least, its squares $v(r_i,h_j)^2 = v(r_i,h_{-j})^2 $). What we simply need to show is:

\begin{equation}
\sum\limits_{n=-N}^{-1} \int_{h_n}^{h_{n+1}} \mathcal{S}^{(i,j)}(\hat{r} , \hat{h}) \hat{\rho}(\hat{r},\hat{h})\hat{r}  \spaceI d\hat{r} d\hat{h} = \sum\limits_{n=0}^{N-1} \int_{h_n}^{h_{n+1}} \mathcal{S}^{(i,j)}(\hat{r} , -\hat{h}) \hat{\rho}(\hat{r},\hat{h})\hat{r}  \spaceI d\hat{r} d\hat{h} \spaceI .
\end{equation}

The derivation of this essential property is briefly developed in appendix \ref{AppendixSymmetricProperty}. Moreover, without any loss of genarility, let us now begin the indexes from 1 in order to be in consonance with other chapters. Thus, the equation \eqref{GoverningEquationsNoSymmetry} now is

\begin{equation}
\sum\limits_{m=1}^{M-1} \sum\limits_{n=1}^{N-1}  \int_{r_m}^{r_{m+1}} \int_{h_n}^{h_{n+1}} \mathcal{G}^{(i,j)}(\hat{r} , \hat{h}) \hat{\rho}(\hat{r},\hat{h}) \spaceI d\hat{r} d\hat{h} + \frac{A}{2} v(r_i,h_j)^2 = 0 \spaceI ,
\end{equation}
where

\begin{equation}\label{Gcaligraphic}
\mathcal{G}^{(i,j)}(\hat{r} , \hat{h}) = \left[ \mathcal{S}^{(i,j)}(\hat{r} , \hat{h}) + \mathcal{S}^{(i,j)}(\hat{r} , -\hat{h}) \right] \hat{r} \spaceI .
\end{equation}

Now we have reduced our system into a $(M+1)\times(N+1)$ equations system and from now on, there is no other way than solving these integrals. One more time, solving this integral analytically is almost impossible, so we must recur to numerical integration.

In order to avoid any possible singularity at node points, we will recur again to gaussian quadrature, but now we have an adapted version for 2D-integrals (check again appendix \ref{AppendixGaussianQuadrature}). The advantages and basic structure are the same because we only use points inside every square of the net that we have already built. The singularities will be precisely at the net nodes. Here we used a $3 \times 3$ internal grid in order to reduce the code calculations. For this purpose we need to parameterize not only the radius and the height, also the values of the density function on that internal grid. Parametrizing the density distribution could be a way of reducing the calculations required for every point of the internal grid or, at least, a way of standardise the number of calculations given any density distribution function. We can summarize it here as:

\begin{itemize}\label{Parameterization3D}
\item $\hat{r}_m(\xi)=(1-\xi)r_m + \xi r_{m+1} \spaceII \mbox{ \& } \spaceII \hat{h}_n(\eta)=(1-\eta)h_n + \eta h_{n+1} \spaceI .$
\item $\hat{\rho}_{m,n}(\xi,\eta) = \rho_{m,n}(1-\xi)(1-\eta) + \rho_{m,n+1}(1-\xi)\eta + \rho_{m+1,n}\xi(1-\eta) + \rho_{m+1,n+1} \xi \eta \spaceI .$
\item $\abs{\pdiffI{ (\hat{r}_m , \hat{h}_n) }{(\xi,\eta)}} = \diffI{\hat{r}_m}{\xi} \diffI{\hat{h}_n}{\eta} =  (r_{m+1} - r_m)(h_{n+1} - h_n) = s_m u_n \spaceI .$
\end{itemize}

After all the substitutions we get a governing equations set of the form:

\begin{equation}\label{GoverningEquation3DParam}
\sum\limits_{m=1}^{M-1} \sum\limits_{n=1}^{N-1}  \int_{0}^{1} \int_{0}^{1} \mathcal{G}_{(m,n)}^{(i,j)}(\xi,\eta) \hat{\rho}_{m,n}(\xi,\eta) s_m u_n  \spaceI d\xi d\eta \spaceI + \spaceI \frac{A}{2} v(r_i,h_j)^2 = 0 \spaceI . 
\end{equation}

Where the function $\mathcal{G}$ fully parameterized can be finally expressed by:

\begin{equation}\label{GcaligraphicParameterized}
\mathcal{G}^{(i,j)}_{(m,n)} = \left[ \mathcal{S}^{(i,j)}( \hat{r}_m , \hat{h}_n ) + \mathcal{S}^{(i,j)}( \hat{r}_m , - \hat{h}_n ) \right] \hat{r}_m = \left[ \mathcal{S}^{(i,j)}_{(m,n^+)} + \mathcal{S}^{(i,j)}_{(m,n^-)} \right] \hat{r}_m \spaceI .
\end{equation}

As we have done in the second chapter, now we have the chance of making similar calculations, this time over the galactic plane. Here, we have got the easy situation. If we suppose to have a theoretical density profile and the value of the galactic parameter we can be able to know the velocity of the rotation curve in every point. From equation \eqref{GoverningEquation3DParam} we can solve for the velocity.

\begin{equation}\label{Curves3D}
v(r_i,h_j) = \left( - \frac{2}{A} \sum\limits_{m=1}^{M-1} \sum\limits_{n=1}^{N-1}  \int_{r_m}^{r_{m+1}} \int_{h_n}^{h_{n+1}} \mathcal{G}^{(i,j)}(\hat{r} , \hat{h}) \hat{\rho}(\hat{r},\hat{h})  \spaceI d\hat{r} d\hat{h} \right)^{1/2} \spaceI .
\end{equation}

If we parametrize it, we get the analogue equation to equation \eqref{Curves2DParam}:

\begin{equation}\label{Curves3DParam}
v(r_i,h_j) = \left( - \frac{2}{A}  \sum\limits_{m=1}^{M-1} \sum\limits_{n=1}^{N-1}  \int_{0}^{1} \int_{0}^{1} \mathcal{G}_{(m,n)}^{(i,j)}(\xi,\eta) \hat{\rho}_{m,n}(\xi,\eta) s_m u_n  \spaceI d\xi d\eta   \right)^{1/2} \spaceI .
\end{equation}

The angular integral reduction \eqref{IntegralReduction2D} is not so necessary here. It have been built with the inverse problem (section \ref{SectionDensityProfileCalculus3D} ) in mind. Moreover, even though the parametrization and the use of gaussian quadrature provides of a solidus integration method, it increments considerably the running time. As we have said, we used $3 \times 3$ internal grid, but we think even a $2 \times 2$ internal would be enough.

\subsection{Flare influence}

As we have have seen in subsection \ref{SubsectionFlaredDisc}, the effect of a flare cannot be measured using the thin disc hypothesis. Furthermore, the equation \eqref{Curves3DParam} offers us a good chance to check how a flare can affect the usual rotation curve of a flared disc.

Here we have not taken into account a whole Galaxy model, just the disc distribution. However, it is a good example to check how the flare can affect to the Galactic rotation curve. For this purpose we will modify the distribution we have already described in \ref{SubsectionFlaredDisc}, composed by the equations \eqref{FlaredDisc}, \eqref{FlaredDiscComponents} and \eqref{FlaredDiscScales}. Those equations do not take special care for the distribution of the sections within the solar radius. So we mix their data and structure with the classical Freeman exponential disc of equation \eqref{ExponentialDiscFreeman} of the Milky Way.

\begin{equation}
\rho(r,h) = \rho_{d,0} \exp\left( - \dfrac{r}{R_d} \right) \dfrac{ h_z(R_{\Sun}) }{ h_z(r) }  \exp \left( - \dfrac{\abs{z}}{h_z(r)} \right) \spaceI .
\end{equation}

We have obtained the central disc density from equation \eqref{ExponentialDiscMass}, and it takes the value $\rho_{d,0} \approx 3$ M$_{\Sun}$ pc$^{-3}$, ten times the solar neighbourhood spatial density. We take the estimated disc mass as $M_d = 5 \times 10^{10}$ M$_{\Sun}$ (McMillan , \cite{McMillan}). The other components are described in subsection \ref{SubsectionFlaredDisc}. The starting point of the flare is the solar radius. We have considered as integration domain the radius till 30 kpc and the height till 7 kpc (a little bit greater than the height scale at $r=30$ kpc). This distribution also preserves the essential property: The distribution does not add any mass to the disc, just spread it through the vertical direction, the equation of the conservation of projected mass, \eqref{FlareProjected}, is also valid here. The plotted velocities corresponds to rotation curves at the galactic plane ($h=0$ kpc) and $h=3$ kpc.

First of all, we can observe in figure \ref{FigureFlares} the most intuitive result: As we get away from the Galactic plane in the vertical axis, the rotation curve gradually reduces its value. That is a clear effect produced by the density decrement.

\begin{figure}[h!] 
\centering
\includegraphics[width=0.7\textwidth]{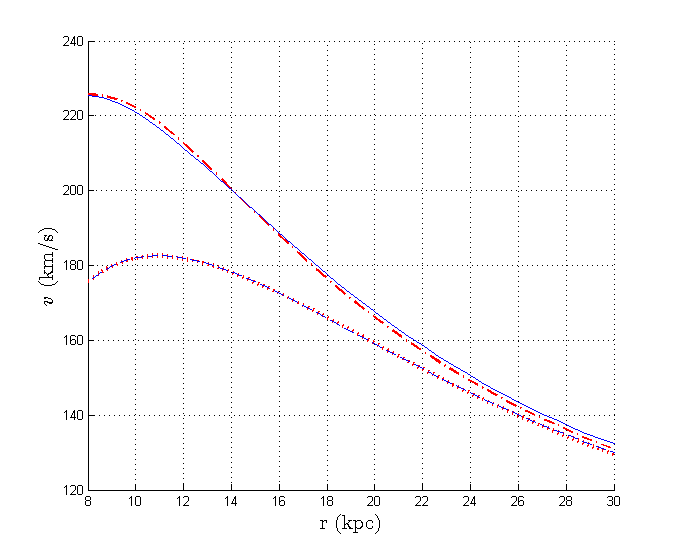}
\caption{Flare effect. No flare: Blue line ($h=0$ kpc) and dashed line ($h=3$ kpc). Flare: Red point-dashed line ($h=0$ kpc) pointed line ($h=3$ kpc).}
\label{FigureFlares}
\end{figure}

The comparison between the influence of flare and non-flared disc in the Galactic plane is not negligible, but it does not affect so much to the original data. However there is some important points to comment. At the point where the flare starts ($R_{\Sun}$) the flare seems to provoke a velocity increase until $r\sim 14$ kpc where the flare effect is the opposite, producing gradually lower velocities, which is the expected result. However the differences between velocities are no greater than the 1 \% in the galactic plane as we can observe in figure \ref{FigureFlaresComparison}. In higher heights the effect is not the same. The case situated at $h=3$ kpc has a much weaker influence of the flare. The difference of velocities looks to be even inverted (where the flare increase the velocity, here the velocity is decreased and vice versa), but beyond $r \sim 22$ kpc looks like the curve has a more regular decrease.  

Notwithstanding the foregoing, we must note the 3D governing equation given from the beginning, \eqref{GoverningEquation3D}, is still an approximation. While we were computing these results we obtained different curves, and we think there is some tendency to produce higher rotation curves than usual ones.

\begin{figure}[h!] 
\centering
\includegraphics[width=0.7\textwidth]{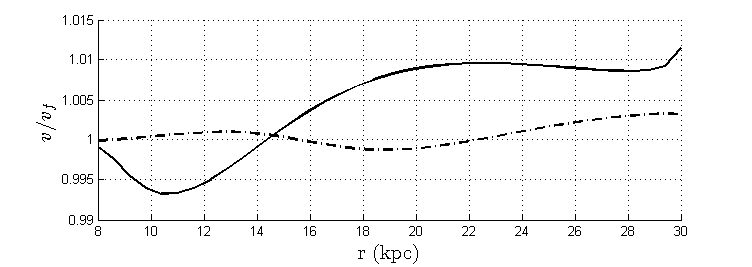}
\caption{Flare comparison in the Galactic plane. Relation ($v/v_f$) between the velocities produced from a non-flared distribution and a flared one. The lines are related to the rotation curves at $h=0$ kpc and the dashed lines are related to the rotation curves at $h=3$ kpc.}
\label{FigureFlaresComparison}
\end{figure}

\section{A sketch of the density profile calculus in 3D}\label{SectionDensityProfileCalculus3D}

Even though we will not work it out here, the idea of reproducing the third chapter in order to get values out of the galactic plane has been considered. We will just sketch here how the system may be structured and give an idea of the path to the solution. Besides, the real support of this numerical problem is the angular integral reduction \eqref{IntegralReduction2D} which let us make the problem easy to compute, saving us from many calculations, complex integrals and so, reducing the running time. Furthermore, the fact of reducing this triple integral becomes a must-do here.

Following the same reasoning that we developed in the third chapter, all the calculations will lead us to a linear system. A huge one in case we want to introduce a thin grid. Similar chapter three explanation should be done here. Observing our density distribution parameterization equations \eqref{Parameterization3D}, we can extract the node values as unknowns variables, $\rho_{m,n}$. With the set of equations \eqref{GoverningEquation3DParam} we can build the linear system we have been seeking from the beginning. Every value of $(i,j)$ corresponds to an equation in the system that also contains all the unknowns $\rho_{m,n}$ and the galactic parameter $A$, which is another unknown. So we would have $M \times N$ equations and $M \times N + 1$ unknowns.

What we will get is another integral equation where the kernel function will be numerically represented by the functions $\mathcal{G}^{(i,j)}_{(m,n)}$ of equation \eqref{GcaligraphicParameterized}. This time, the elements of the linear system matrix $\Gamma$ will be composed of four elements similar to the ones given in \eqref{GammaSubElements}; one for every corner of every square of the discretization grid. That is, very element of the matrix will depend of the four indexes, $\Gamma^{(i,j)}_{(m,n)}$. We can sketch the structure of the $\Gamma$ matrix and the order of its indexes as:

\begin{equation}
\begin{array}{cc|c|c|c|c|} \cline{3-6}
      	& j=1 \rightarrow &  				 	&  				 	&		 &  \\
 i=1  	& \vdots			  & m=1 ; \forall n 	& m=2 ; \forall n 	& \cdots & v(r_1,h_j)^2/2 \\
      	& j=N \rightarrow &  				 	&  				 	&        &  \\ \cline{3-6}

      	& j=1 \rightarrow &  				 	&  				 	&        &  \\
 i=2  	& \vdots 		  & m=1 ; \forall n 	& m=2 ; \forall n 	& \cdots & v(r_2,h_j)^2/2 \\
      	& j=N \rightarrow &  				 	&  				 	&        &  \\ \cline{3-6}

\vdots& \vdots & m=1 ; \forall n 	& m=2 ; \forall n 	& \cdots & v(r_i,h_j)^2/2 \\ \cline{3-6}

      	& j=1 \rightarrow &  				 	&  				 	&        &  \\
 i=M		& \vdots			  & m=1 ; \forall n 	& m=2 ; \forall n 	& \cdots & v(r_M,h_N)^2/2 \\
      	& j=N \rightarrow &  				 	&  				 	&        &  \\ \cline{3-6}

CME  & \longrightarrow & m=1 ; \forall n 	& m=2 ; \forall n 	& \cdots & 0 \\ \cline{3-6}

\end{array}
\end{equation}   

In order to have one more equation and some independent coefficient in the equations we have added the conservation mass equation with the symmetric property of the density, \eqref{MCE}. Furthemore we will be able to calculate the GRN, $A$. If we parameterize it through equations \eqref{Parameterization3D} we will get

\begin{equation}
\sum\limits_{m=1}^{M-1} \sum\limits_{n=1}^{N-1} \int_{r_m}^{r_{m+1}} \int_{h_n}^{h_{n+1}} \hat{\rho}(\hat{r},\hat{h}) \hat{r} \spaceI d\hat{r} d\hat{h} = \dfrac{1}{4 \pi}
\end{equation}																							 						 									
Now, we would ``only'' have to take care of the boundaries, find a way about how to manage the possible instabilities, and maybe find another way of reducing the calculation time without compromising the robustness of the numerical integration.

% ==============================================================================================
% CHAPTER ======================================================================================
% ==============================================================================================

\chapter{Conclusions}

We have arrived to this point by departing from a specific method proposed in {F\&G} article \cite{FengGallo2011}. This method has been used in two directions: one of them enabled us to get the rotation curve of a galaxy if we already knew its density distribution throughout the disc. The `way back' (what we have called the `inversion problem') is a little more complicated and requires a deeper understanding of what we are managing. The hypothesis of a thin disc is required to get appropriate results and it is very important to emphasize that this method interprets the gravitational system as self-consistent. That is, all the gravitational mass is enclosed within the thin disc and there is no matter outside this boundary. In this thin disc the vertical distribution of the mass is considered constant along the thin vertical distance. Furthermore, in extracting our conclusions, we must look to the physical consequences when we think of the output produced.

The derivation of rotation curves given a density distribution does not seem to produce any problem. The rotation curve produced fits with established theoretical and analytical models. The Freeman-Mestel distribution is presented as a good alternative to the Mestel distribution in the case where we wish to avoid the singularity contained in the Mestel distribution equation \eqref{MestelCurve}. At the same time it will preserve the main property of being able to generate a rotation curve similar to those produced for spiral galaxies, with all the galactic mass within the disc. Despite of some differences between the computed curve and the Mestel theoretical curve in the regions closest to the galactic centre, the computed curve has a reasonable shape. At the same time, it confirms the clear difference of meaning of the core radius for every distribution. However, both distributions tend to the same curve if we make the core radius tend to zero (figure \ref{FigureCurvesMestel}).

At the same time, following Binney \& Tremaine's comparison in \ref{Binney} (Fig. 2-17), we can check how different the rotation curves produced for the same amount of matter shared in different distributions can be. If we concentrate the mass in a thin disc, the rotation curves produced will be greater than other more expanded, e.g.\ spherical, distributions. However, as we get close to the periphery, all the curves will gradually tend to the Keplerian descending curve (see figure \ref{FigureExponentialCurves}).

In all these examples and test, the Galactic Rotation Number (GRN) must be part of the input. The theoretical value for Mestel's distribution ($A=\pi/2$) has been used in subsection \ref{SubsectionFMcurves} (where the full galaxy mass must be considered). But in the case of the last paragraph, the data introduced is only related to the full disc. Furthermore, the output data are clearly plausible and realistic, which is another good sign. Finally, the flared situation gives us no clear results, since the projected mass distribution offers no difference with the non-flared projected mass.

In the third chapter, the obstacles we find can offer us a good set of ideas to write here. First of all, we must underscore that we have reached similar results using different numerical methods, as we have explained in section \ref{SectionStabilityProblem}. Integral equation and inverse problem are dealt with there. What was really disturbing about our instabilities was precisely how we found the same structural obstacle through two ways of solving the same problem. It is not easy to improve a given solution in an integral equation.

However, if we modify the original data and select appropriate parameters in a careful and moderate way, we can get acceptable results as we do in section \ref{SectionMilkyWayDensityDistribution}. Moreover, this approach is a good for checking the consistency of the governing equation \eqref{GoverningEquation2D}, and so the {F\&G} proposal regarding Newtonian dynamics as a good way to tackle this problem, in spite of the natural singularity we face when we are dealing with Newton's formulation.

One of the interesting points about the distributions obtained from theoretical curves with a specific galactic core is the GRN obtained. This number connects the value of the entire galactic mass and the radius. {F\&G} concluded in \cite{FengGallo2011} that, for most of the SB-type galaxies with rotation curves similar to that of the Milky Way galaxy, we get $A\approx1.6$. If we check again the equation \eqref{A}, we can easily see how this result led them to the conclusion of the proportionality of the galactic mass, $M_g$, and the disc size, $R_g$. It is very important to emphasize here again that this method interprets the gravitational system as self-consistent. That is, if we input a rotation curve to some radius, the program will think there is no matter outside the disc. This exerts a clear influence on the periphery of the distribution, making the distribution values gradually fall to zero throughout a smooth curve. The finite size implies a finite amount of mass in Newtonian dynamics.

The comparison between our computed densities of the Milky Way in section \ref{SectionMilkyWayDensityDistribution} (see figure \ref{FigureMWdistribution}) and those obtained by {F\&G} in \cite{FengGallo2011} (Fig. 7) and \cite{FengGallo2014} (Fig. 3) are not so different. Since we have carefully modified the bulge, the first parsecs just before the usual flat central density distribution show a slightly different initial descending curve from the galactic core. Of course, the wiggles produced for different ripples in the original different data we have used are not exactly the same. But the point here is that we have taken a greater radius. In their articles F\&G usually use 16 kpc $\lesssim R_g \lesssim$ 20 kpc. For the most accurate curve we have used $R_g = 24$ kpc. The surface mass density we obtained in the solar neighbourhood ($1.06 \times 10^2$ M$_{\Sun}$ pc$^{-2}$) is a little behind the {F\&G} value ($1.44 \times 10^2$ M$_{\Sun}$ pc$^{-2}$). Both of them are closer to the bulge+disc data ($0.89 \times 10^2$ M$_{\Sun}$ pc$^{-2}$) than the bulge+disc+halo data ($4.2 \times 10^2$ M$_{\Sun}$ pc$^{-2}$). This confirms that the mass enclosed within the disc really has a higher influence on rotation curves than that great vertical distances from the Galactic plane. Perhaps the increment of the Galactic disc size (and so, the disc mass in the periphery) could contribute to increment the surface mass density required to hold the rotation curves. If we computed the same rotation curve, with a considerably smaller galaxy radius ($R_g=16$ kpc)  and the same thickness of the disc ($H=240$ pc), we would get a more accurate Surface Mass Density (SMD) in the solar neighbourhood ($0.88 \times 10^2$ M$_{\Sun}$ pc$^{-2}$). However, this would need a more detailed analysis.

In the last paragraph we have been speaking about de 2013 Sofue data. But the 2012 data that we also fitted is a consequence of the inaccuracy of the measurements taken for radius beyond 10 kpc. Calculating how the errors and standard deviation are transmitted through the code would be a good idea to continue on from this essay. And these data would be a perfect base to work on it.

The decomposition of the Galactic rotation curve into the de Vaucouleurs bulge, exponential disc, and isothermal halo may not fit perfectly with the 2013 Sofue data, but it is good enough for our modest target here. {F\&G} also carry out some different decompositions in \cite{FengGallo2014}. One of these just divides the curve rotation into the galactic core and a massive thin disc. Another shows a disc with a prominent cusp in the core and an extended spherical bulge that causes the upper wiggle that we can usually see between $\sim$2.5 kpc and $\sim$8 kpc, which is usually attributed to the exponential disc. Their proposals may not be common ones, so we decided to use the most classical one to estimate the dark disc.

Subtituting the dark halo by a dark disc topic can be quite controversial here. {F\&G}, by computing some solutions with a `cut-off' radius at 20.55 kpc, obtain a full galaxy mass sufficient to support the Milky Way disc rotation \cite{FengGallo2014}. Following this reasoning and handling some density data in \cite{FengGalloDarkMatter}, they estimated a gas density of the interstellar medium (ISM) and finally conclude that `the amount of mass required to support the observed rotation curve could be no more than that found in typical ISM.' Moreover, they have provided further justification in recent papers about $N$-Body simulations: `a disc galaxy with an almost flat rotation curve can be stabilized by dense centers without the dark matter halo' (see Sellwood \& Evans in \cite{Sellwood2001}).

The dark disc we have estimated has a clear dominance beyond 11--12 kpc. As we have commented in section \ref{SectionDarkDisc}, the required density to hold the rotation disc within the thin disc has a maximum of $\Delta \rho \lesssim 0.1$ M$_{\Sun}$ pc$^{-3}$ around $r\sim$ 15--16 kpc. These value correspond in terms of SMD to $\Delta \Sigma \lesssim 24$ M$_{\Sun}$ pc$^{-2}$. A recent paper about dark disc constraints (Kramer \& Randall in \cite{KramerRandall2016}) stablishes that `the kinematics currently allow for a thin dark disc of up $\Delta \Sigma \lesssim 14$ M$_{\Sun}$ pc$^{-2}$ of SMD, with a weaker bound for thicker discs.' The result we have obtained is clearly above that value and not so believable for just a baryonic component.

Finally, the chapter dedicated to the 3D extension left us interesting results. The velocity at large radii seems not to be so affected by the flare effect. Although the distribution used does not add any mass to the disc, it is easy to suppose that a small increment of mass on flare distances could slightly aid rotation curve support. In any case, the flare mainly provokes a faster velocity decrease in the outer part of the Galaxy.

%==================================================================================================
% APPENDIX  =======================================================================================
%==================================================================================================

\begin{appendices}
\chapter{}

\section{Elliptic Integrals}\label{AppendixEllipticIntegrals}

The incomplete elliptic integrals of first and second kind are:

\begin{equation}\label{EllipticIntegrals}
F( \theta, k) = \int_0^{\theta} \dfrac{d \phi}{\sqrt{ 1 - k^2 \sin^2 \phi }}; \spaceII E( \theta, k) = \int_0^{\theta} \sqrt{ 1 - k^2 \sin^2 \phi } \spaceI d\phi \spaceII ,
\end{equation}
where $k \in [0,1]$ is the \textit{elliptic modulus} or \textit{eccentricity}. In many other references it is common to use de \textit{elliptic parameter}, $m$. The relation between them is $m=k^2$. The complete elliptic integrals are the ones with $\theta=\pi/2$ and we can write them as their \textit{Jacobi's form} just by substituting $t=\sin \phi$.

\begin{equation}\label{EllipticIntegralsComplete}
K(k) \equiv F \left( \frac{\pi}{2} , k \right) = \int_0^1 \dfrac{dt}{\sqrt{ (1-t^2)(1 - k^2 t^2) }}; \spaceII E(k) \equiv E \left( \frac{\pi}{2},k \right) = \int_0^1 \sqrt{ \dfrac{1 - k^2 t^2}{1-t^2} } \spaceI dt \spaceI .
\end{equation}

%-----------------------------------------------------------------------------------------------------------------------
\section{Condition Number}\label{AppendixConditionNumber}
 If we have a linear system of the form $Au=b$, and we note the data perturbation $b + \delta b$ and its corresponding perturbed solution $u + \delta u$, we define the \textit{condition number} of the system/matrix asociated to a p-norm as:

\begin{equation}
\mbox{cond}_p(A) = \norm{A}_p \norm{A^{-1}}_p.
\end{equation}

Independently of $p$, it satisfies

\begin{equation}
\dfrac{\abs{\delta u}}{\abs{u}} \leq \mbox{cond}(A) \dfrac{\abs{\delta b}}{\abs{b}} \spaceII \mbox{\&} \spaceII \dfrac{\abs{\delta u}}{\abs{u + \delta u}} \leq \mbox{cond}(A) \dfrac{\norm{\delta A}}{\norm{A}}.
\end{equation}

And its the second inequality whom show us the idea of condition number as a magnification factor for our problem: The supposed inaccuracy in numerical representation of the kernel function. The idea of increasing the nodal points to reduce the discretisation errors is another dead end. We have that:

\begin{equation}
\convergence{\mbox{cond}(A)}{\infty}{n}{\infty}.
\end{equation}

In practice you can observe this when you compute the system with large discretisation sets. There is more information about condition number and stability in literature \cite{Infante} \cite{CraigBrawn}.

%-----------------------------------------------------------------------------------------------------------------------
\section{Solved Integrals}\label{AppendixSolvedIntegrals}

These two solved integrals can be found in \textit{Gradshteyn \& I.M. Ryzhik's} (\cite{Gradshteyn}, pages 179 \& 182) or \textit{P.F. Byrd \& M.D. Friedman} (\cite{Friedman}, pages 176 \& 177 ). The second reference is an amazingly complete handbook just dedicated to the elliptic integrals. However we will use the simple notation of \textit{Gradshteyn \& I.M. Ryzhik's}. The solved integrals are

\begin{eqnarray}
I_1 =  \int_0^{\Phi} \dfrac{ d \phi }{ ( a - b\cos \phi )^{1/2} }  & = & \dfrac{2}{\sqrt{ a+b }} F(\delta,k) \\
I_3 = \int_0^{\Phi} \dfrac{ d \phi }{ ( a - b\cos \phi )^{3/2} } & = & \dfrac{2}{ (a-b) \sqrt{a+b} } E(\delta,k) \spaceI ,
\end{eqnarray}

where

$$ \Phi \in [0,\pi]; \spaceII \sin \delta = \sqrt{ \dfrac{(a+b)(1-\cos \Phi)}{ 2( a - b \cos \Phi ) } }; \spaceII k=\sqrt{\dfrac{2b}{a+b}}; \spaceII  a>b>0 \spaceI .$$

Knowing our values ( $a = \hat{r}^2 + r^2 + \Delta h^2; \spaceI b = 2\hat{r}r$ ), there is no difficulty on showing that

$$ \delta \in [0,\pi/2]; \spaceII k^2 = \frac{4\hat{r}r}{ (\hat{r}+r)^2 + \Delta h^2}; \spaceII \hat{r}^2 + r^2  + \Delta h^2 > 2\hat{r}r >0 \spaceI .$$

%-----------------------------------------------------------------------------------------------------------------------
\section{Symmetric Property of the disc}\label{AppendixSymmetricProperty}

First of all we simply do $\{n \rightarrow (-n-1) \}$. Then

\begin{eqnarray}
\sum\limits_{n=-N}^{-1} \int_{h_n}^{h_{n+1}} \mathcal{S}_{(i,j)}(\hat{r} , \hat{h}) \hat{\rho}(\hat{r},\hat{h})\hat{r}  \spaceI d\hat{r} d\hat{h} & = & \sum\limits_{n=0}^{N-1} \int_{h_{-n-1}}^{h_{-n}} \mathcal{S}_{(i,j)}(\hat{r} , \hat{h}) \hat{\rho}(\hat{r},\hat{h}) \hat{r} \spaceI d\hat{r} d\hat{h} \nonumber \\
\{ h_j=-h_{-j} \} \rightarrow & = & \sum\limits_{n=0}^{N-1} \int_{-h_{n+1}}^{-h_{n}} \mathcal{S}_{(i,j)}(\hat{r} , \hat{h}) \hat{\rho}(\hat{r},\hat{h}) \hat{r} \spaceI d\hat{r} d\hat{h}  \nonumber \\ 
\{ \tilde{h} = -\hat{h} \} \rightarrow  & = & \sum\limits_{n=0}^{N-1} \int_{h_n}^{h_{n+1}} \mathcal{S}_{(i,j)}(\hat{r} , -\tilde{h}) \hat{\rho}(\hat{r},-\tilde{h})\hat{r}  \spaceI d\hat{r} d\tilde{h} \nonumber \\
\{ \hat{\rho}(\hat{r},\tilde{h}) = \hat{\rho}(\hat{r},-\tilde{h}) \} \rightarrow  & = & \sum\limits_{n=0}^{N-1} \int_{h_n}^{h_{n+1}} \mathcal{S}_{(i,j)}(\hat{r} , -\tilde{h}) \hat{\rho}(\hat{r},\tilde{h})\hat{r}  \spaceI d\hat{r} d\tilde{h} \spaceI . \nonumber \\
\end{eqnarray}

%-----------------------------------------------------------------------------------------------------------------------
\section{Gaussian quadrature}\label{AppendixGaussianQuadrature}

	\begin{enumerate}
	
	\item Case for 1D
	
	\begin{equation}
	\int_a^b f(x) dx = \dfrac{b-a}{2}  \int_{-1}^1 f\left(  \dfrac{b-a}{2} x + \dfrac{b+a}{2}  \right) dx \approx \dfrac{b-a}{2} \sum_{i=1}^n w_i f\left(  \dfrac{b-a}{2} x_i + \dfrac{b+a}{2}  \right).
	\end{equation}
	
	In our case we have $a=0$ and $b=1$. Thus
	
	\begin{equation}
	\int_0^1 f(x) dx \approx \dfrac{1}{2} 	\sum_{i=1}^n w_i f\left( \dfrac{1}{2} (x_i+1)  \right) \spaceI ,
	\end{equation}
	
	where $ x_i \in [-1,1] $. And the weights $w_i$ are defined here by Legendre Polynomials. From \cite{AbramowitzStegun1972} we can take

	\begin{equation}
	w_i=\dfrac{2}{  (1-x_i^2)[  P'_n (x_i)  ]^2  }  \spaceI .
	\end{equation}		
	
	We will use the 6-points quadrature, so we get as weight functions:
	
\begin{table}[h!]
\centering
	\begin{tabular}{|c|c|c|c|c|} \hline
$i$ & Weight $w_i$ & $x_i \in [-1,1]$ & $x_i \in [0,1]$ \\	\hline
$1$ & $ 0.1713244923791704 $ & $-0.9324695142031521 $ & $ 0.033765242898424 $ \\	\hline
$2$ & $ 0.3607615730481386 $ & $-0.6612093864662645 $ & $ 0.169395306766868 $ \\	\hline
$3$ & $ 0.4679139345726910 $ & $-0.2386191860831969 $ & $ 0.380690406958402 $ \\	\hline
$4$ & $ 0.4679139345726910 $ & $ 0.2386191860831969 $ & $ 0.619309593041598 $ \\	\hline
$5$ & $ 0.3607615730481386 $ & $ 0.6612093864662645 $ & $ 0.830604693233132 $ \\	\hline
$6$ & $ 0.1713244923791704 $ & $ 0.9324695142031521 $ & $ 0.966234757101576 $ \\	\hline
	\end{tabular}
\end{table}

	\item The bidimensional case has a similar look:
	
	\begin{equation}
	\int_a^b \int_a^b f(x,y) dx dy  \approx \left( \dfrac{b-a}{2} \right)^2  \sum_{i=1}^n \sum_{j=1}^n  w_i w_j  f\left(  \dfrac{b-a}{2} x_i + \dfrac{b+a}{2} , \dfrac{b-a}{2} y_j + \dfrac{b+a}{2}  \right) \spaceI .
	\end{equation}
	
	In our integration interval we finally get
	
	\begin{equation}
	\int_0^1 \int_0^1 f(x,y) dx dy \approx \dfrac{1}{4} \sum_{i=1}^n \sum_{j=1}^n w_i w_j f \left(  \dfrac{1}{2} (x_i+1) , \dfrac{1}{2} (y_j+1)  \right) \spaceI .
	\end{equation}
	
	The table for 3 points quadrature contains the following values.
	
\begin{table}[h!]
\centering
	\begin{tabular}{|c|c|c|c|c|} \hline
$i$ & Weight $w_i$ & $x_i \in [-1,1]$ & $x_i \in [0,1]$ \\	\hline
$1$ & $ 0.555... $ & $-0.7745966692414834 $ & $ 0.112701665379258 $ \\	\hline
$2$ & $ 0.888... $ & $ 0					 $ & $ 0.5 				 $ \\	\hline
$3$ & $ 0.555... $ & $ 0.7745966692414834 $ & $ 0.887298334620742 $ \\	\hline
	\end{tabular}
\end{table}

\end{enumerate}

\end{appendices}

% ==================================================================================
% Bibliography =====================================================================
% ==================================================================================

%===========================================================================================================================================================
\end{document}